\documentclass[fleqn,usenatbib]{mnras}

\usepackage{newtxtext,newtxmath}

\usepackage[T1]{fontenc}
\usepackage{ae,aecompl}

\usepackage{graphicx}	
\usepackage{amsmath}	
\usepackage{amssymb}	
\usepackage[dvipsnames]{xcolor}

\graphicspath{ {figures/} }

\newcommand{\HI}{\ion{H}{i}}
\newcommand{\hkpc}{h^{-1}{\rm kpc}}
\newcommand{\hmpc}{h^{-1}{\rm Mpc}}
\newcommand{\gad}{{\sc Gadget-3}}
\newcommand{\gizmo}{{\sc Gizmo}}
\newcommand{\mufasa}{{\sc Mufasa}}
\newcommand{\simba}{{\sc Simba}}
\newcommand{\fmol}{f_{\rm H2}}
\newcommand{\tH}{t_{\rm H}}
\newcommand{\dmsq}{\Delta_{\rm MSQ}}

\definecolor{mycolor}{rgb}{0,0,0}

\title[Mergers, starbursts, and quenching in SIMBA]{Mergers, Starbursts, and Quenching in the Simba Simulation}

\author[Rodr\'iguez et al.]{Francisco Rodr\'iguez Montero$^{1}$\thanks{E-mail: \textcolor{mycolor}{currodri@gmail.com}},
Romeel Dav\'e$^{1,2,3}$,
Vivienne Wild$^4$,
\newauthor
Daniel Angl\'es-Alc\'azar$^{5}$, 
Desika Narayanan$^{6,7,8}$
\\
\\$^1$ Institute for Astronomy, Royal Observatory, Edinburgh EH9 3HJ, United Kingdom
\\$^{2}$ University of the Western Cape, Bellville, Cape Town 7535, South Africa
\\$^{3}$ South African Astronomical Observatories, Observatory, Cape Town 7925, South Africa
\\$^{4}$ School of Physics \& Astronomy, University of St Andrews, North Haugh, St Andrews, Fife KY16 9SS, UK
\\$^5$ Center for Computational Astrophysics, Flatiron Institute, 162 Fifth Avenue, New York, NY 10010, USA
\\$^6$ Department of Astronomy, University of Florida, 211 Bryant Space Sciences Center, Gainesville, FL, USA
\\$^7$ University of Florida Informatics Institute, 432 Newell Drive, CISE Bldg E251, Gainesville, FL, USA
\\$^8$ Cosmic Dawn Center at the Niels Bohr Institute, University of Copenhagen\
 and DTU-Space, Technical University of Denmark
}

\date{Accepted XXX. Received YYY; in original form ZZZ}

\pubyear{2019}

\begin{document}
\label{firstpage}
\pagerange{\pageref{firstpage}--\pageref{lastpage}}
\maketitle

\begin{abstract}
We use the \simba\ cosmological galaxy formation simulation to investigate the relationship between major mergers ($\la$ 4:1), starbursts, and galaxy quenching.  Mergers are identified via sudden jumps in stellar mass $M_*$ well above that expected from in situ star formation, while quenching is defined as going from specific star formation rate sSFR$>\tH^{-1}$ to sSFR$<0.2\tH^{-1}$, where $\tH$ is the Hubble time.  At $z\approx 0-3$, mergers show $\sim\times 2-3$ higher SFR than a mass-matched sample of star-forming galaxies, but globally represent $\la 1\%$ of the cosmic SF budget.  At low masses, the increase in SFR in mergers is mostly attributed to an increase in the $H_2$ content, but for $M_*\ga 10^{10.5} M_{\sun}$ mergers also show an elevated star formation efficiency suggesting denser gas within merging galaxies. The merger rate for star-forming galaxies shows a rapid increase with redshift $\propto (1+z)^{3.5}$, but the quenching rate evolves much more slowly, $\propto (1+z)^{0.9}$; there are insufficient mergers to explain the quenching rate at $z\la 1.5$.  \simba\ first quenches galaxies at $z\ga 3$, with a number density in good agreement with observations. The quenching timescales $\tau_q$ are strongly bimodal, with `slow' quenchings ($\tau_q\sim 0.1\tH$) dominating overall, but `fast' quenchings ($\tau_q\sim 0.01t_H$) dominating in $M_*\sim 10^{10}-10^{10.5}M_{\sun}$ galaxies, likely induced by \simba's jet-mode black hole feedback.  The delay time distribution between mergers and quenching events suggests no physical connection to either fast or slow quenching.  Hence, \simba\ predicts that major mergers induce starbursts, but are unrelated to quenching in either fast or slow mode.
\end{abstract}




\section{Introduction}\label{sec:intro}
The increasing sample size and precision of present-day deep sky surveys have transformed our view about the evolution of galaxies over cosmic time. With such surveys characterising the statistical properties of galaxies to great precision, theoretical models about the large scale assembly of cosmological structures can be rigorously tested. Models of galaxy formation such as semi-analytical models~\citep{Benson_2010} and hydrodynamical simulations~\citep{Somerville_2015} must include numerous small-scale processes related to star formation, black hole growth, and energetic feedback processes in order to reproduce the observed galaxy population. Despite the progress achieved in the last decades, we still lack a clear understanding of the physics driving these so-called sub-grid processes in large-scale models of cosmological galaxy formation.

\textcolor{mycolor}{A longstanding puzzle is a clear  bimodality  in  galaxy colours, now well} quantified thanks to surveys like the Sloan Digital
Sky Survey~\citep[SDSS;][]{Baldry_2004} and the Galaxy and Mass Assembly Survey \citep[GAMA;][]{Liske_2015}. Observations have shown that the number of stars living in red quiescent galaxies has increased since redshift close to unity, while the stellar mass density in blue star-forming galaxies has remained roughly the same~\citep{Bell_2004, Ilbert_2013, Muzzin_2013}. Hence, the cause of this build up of mass in the red sequence must owe to  the  shutoff  of  star-formation  in  galaxies  and  their  subsequent  migration into  the quiescent population~\citep{Arnouts_2007,Faber_2007}, which is referred to as quenching. A key question in astronomy today is, what physical mechanism(s) give rise to quenching?

Various processes have been proposed to explain galaxy quenching. The proposed solutions can be classified into four physical scenarios:
\begin{enumerate}
    \item \textbf{Halo preventive feedback:} At sufficiently large halo masses, the central cooling time of the hot gas exceeds the Hubble time \citep{Rees_1977}.  Over time, this will starve the galaxy of cold gas, and quench star formation.  However, hierarchically growing halos generally undergo far too much cooling and stellar growth before this halo mass threshold is reached~\citep[e.g.][]{WhiteFrenk_1991}, a problem known as overcooling, and hence this cannot be the sole actor for quenching.
    
    \item \textbf{Maintenance mode feedback:} To counteract the expected cooling in massive halos, some energy source is required.  As massive halos contain little star formation, it is canonical to invoke feedback energy from black hole accretion via active galactic nuclei \citep[AGN; e.g.][]{Bower_2006,Croton_2006,Somerville_2008}.  Evidence from intracluster medium bubbles suggests that AGN jets can provide approximately the required amount of energy to offset cooling~\citep{McNamara_2007}.  

    \item \textbf{Outflow quenching:} AGN are also seen to drive strong molecular and ionised gas outflows~\citep[e.g.][]{Sturm_2011,Diamond_Stanic_2012,Perna_2017}. Estimates of the mass outflow rates suggest that, if those rates continued for a galaxy dynamical time, they could remove all cold gas in the galaxy, thereby quenching star formation. Simulations of gas-rich galaxy mergers that drive starbursts and AGN feedback suggest that the remnant of such a merger would be an elliptical with little cold gas, as observed~\citep{DiMatteo_2005}, and observations of post-starburst galaxies support this scenario~\citep[e.g.][]{Wild_2009}.  
    
    \item {\bf Morphological quenching:} As disk galaxies settle over time~\citep{Kassin_2012}, their gas can stabilise particularly owing to the growth of a dominant bulge component, thereby lowering their star formation efficiency \citep{Martig_2009}.  While this process is unlikely to produce cold gas-free quenched galaxies on its own because it does not remove cold gas, it may aid with the transition \textcolor{mycolor}{or maintenance of} quenching at later epochs.
    
\end{enumerate}
While all of these mechanisms are likely to play a role in quenching at some level, there is not yet a clear consensus about the relative importance of each of these processes. In particularly, while it appears that AGN feedback in some form is required to fully quench galaxies as observed~\citep{Somerville_2015}, the detailed physical mechanisms that tie the feedback energy to the cessation of star formation remain uncertain~\citep{Naab_Ostriker_2017}.

One popular scenario has emerged from the observational connection in the nearby Universe between merger driven Ultra-luminous infra-red starbursts and AGN~\citep{Sanders_Mirabel_1996, Yuan_2010,Alexander_2012}. In this merger-driven quenching scenario, a merger between two star-forming disk galaxies drives a strong circum-nuclear starburst which simultaneously feeds an AGN, resulting in rapid heating and removal of the cold gas to leave a quenched elliptical galaxy with a large black hole. This scenario directly connects mergers, starbursts, black hole growth and quenching, and results in quenching accompanied by morphological transformation that simultaneously explains the star formation properties seen across the Hubble sequence.

The merger-driven quenching scenario has been extensively explored using galaxy formation models.  Early works using gas dynamical simulations showed that galaxy mergers could induce starbursts that used up gas quickly and enacted morphological transformation~\citep{Mihos_1996}. \citet{Springel_2005b} pioneered the inclusion of black hole growth and associated feedback in such isolated merger simulations, and showed that the energy release from the AGN (assumed to be released thermally and spherically) was sufficient to unbind any remaining cold gas and leave a quenched elliptical galaxy.  Analytic models by \citet{Hopkins_2006} and \citet{Hopkins_2009}, based on a suite of isolated merger simulations~\citep{Robertson_2006,Cox_2007}, incorporated this scenario into a cosmological framework, and showed that merger-driven quenching by AGN could plausibly explain numerous observations for the co-evolution of black holes and galaxies.

However, the simulations that were the basis of the merger-driven quenching scenario were not cosmologically situated. Pioneering work by \citet{Sijacki_2007} showed that a similar physical model of black hole growth and feedback implemented in cosmological simulations could co-grow galaxies and black holes as observed. However, it did not yield a quenched galaxy population in full agreement with observations.  This turned out to be quite a challenging endeavour: many cosmological simulations included black hole growth and feedback in various forms, but only in recent years have such simulations been able to simultaneously reproduce the properties of the quenched galaxies and the co-growth of galaxies and black holes. Examples include Illustris~\citep{Vogelsberger_2014}, EAGLE~\citep{Schaye_2015}, Illustris-TNG~\citep{Pillepich_2018}, FABLE~\citep{Henden_2018}, and \simba~\citep{Dave_2019}.  

Meanwhile, semi-analytic models (SAMs) of galaxy formation were also introducing AGN feedback via analytic prescriptions. It was realised early on that while mergers could evacuate gas from galaxies, accretion would eventually restart and return the galaxy to the star-forming population. Motivated for instance by observations of intracluster gas heating from AGN jets~\citep{McNamara_2007},  \citet{Croton_2006} and \citet{Bower_2006} introduced the idea of maintenance mode\footnote{This is alternatively called radio mode feedback, owing to its association with radio jets.} feedback from AGN, which serves to counteract cooling from halo gas.  This eventually starves the central galaxy of cold gas, leading to quenching.  \citet{Gabor_2010} and \citet{Gabor_2012} introduced this scenario into cosmological simulations in a heuristic way, and showed that maintenance mode was necessary to yield long-term quenched galaxies as observed, while merger quenching only temporarily quenched galaxies.  \citet{Gabor_2015} further showed that it can also enact environmental quenching as observed e.g. by \citet{Peng_2010}, in which quenching does not depend on stellar mass but on the galaxy surroundings.  The \mufasa\ simulations~\citep{Dave_2016} employed this approach to yield a quenched galaxy population in very good agreement with observations~\citep{Dave_2017}.

While it appears that maintenance mode AGN feedback is required to produce a stable population of quenched galaxies, it is nonetheless the case that mergers certainly happen, and are seen to eject substantial cold gas. Importantly, mergers may be required to enact the morphological transformation aspect of quenching, although the extent of morphological transformation and strength of quenching depends on the gas content, mass ratios and orbital parameters of the initial galaxies~\citep{Robertson_2006,Hopkins_2009,Johansson_2009}. Recent work by \cite{Martin_2018} using the Horizon-AGN cosmological simulation showed that morphological transformation is enacted by $>1:10$ mergers, but that $>1:4$ major mergers alone are insufficient to explain the majority of today's spheroidal systems \citep[see also][]{Kaviraj_2014}. 

Advancing observations, particularly employing the spatial resolution afforded by the Hubble Space Telescope, have allowed the study of mergers over cosmic time. Owing to the difficulty identifying mergers and quantifying visibility timescales of observational features, there has yet to emerge a clear picture for the cosmic evolution of mergers and their impact on star formation and black hole growth. The merger fraction is generally measured to evolve with redshift, although by how much is still debated \citep[e.g.][]{Bertone_2009,Lotz_2011,Bluck_2012,Duncan_2019}. \citet{Abruzzo_2018} showed using zoom simulations that non-parametric merger indicators at high-$z$ may be difficult to interpret into a merger rate.  A substantial enhancement in star formation is detected in local mergers and close galaxy pairs \citep[e.g.][]{Ellison_2008}, and galaxies at higher redshifts with the highest specific SFR are often more disturbed \citep[e.g.][]{Wuyts_2011}. Nonetheless, the current consensus is that major merger-induced star formation does not dominate the cosmic star formation rate density at any redshift probed \citep[e.g.][]{Jogee_2009,Rodighiero_2011,Kaviraj_2014,Lofthouse_2017}.


The link between mergers and AGN activity is another area of mixed observational results, with different selection and analysis methods yielding apparently inconsistent results. While some authors find an excess of AGN in galaxy pairs and mergers \citep{Ellison_2011} and enhancement of morphological disturbance in AGN \citep{Ellison_2019}, others find that the host galaxies of AGN are generally not more morphologically disturbed than non-AGN hosts \citep{Schawinski_2011,Kocevski_2012, Villforth_2014,Hewlett_2017}.

The link between mergers and quenching is more difficult to tackle directly with observations, due to the time lag between the quenching of star formation and any event that may have caused it.  A curious and potential smoking gun population of galaxies are so-called `post starburst' (PSB), or alternatively `E+A' or `k+a', galaxies.  Characterised by an abundance of A/F stars but a lack of O/B stars, they represent galaxies that have rapidly quenched their star formation, from an initially high sSFR, in the last $\sim$Gyr \citep[e.g.][]{Pawlik_2019}. PSBs in the field are commonly associated with morphological disturbance and tidal features \cite[e.g.][]{Zabludof_1996,Pawlik_2018}, suggesting the influence of strong galaxy-galaxy gravitational interactions and mergers in the quenching process and the nature of PSBs. They contain a high fraction of AGN \citep{Wild_2010}, and their observed properties are largely consistent with them evolving into spheroidal galaxies in several hundred Myr \citep{Yang_2006, Pawlik_2016}. Differences are found for cluster PSBs, where environmental effects such as ram-pressure stripping and weak galaxy-galaxy interactions could also be responsible for the recent shut off in star formation \citep{Aragon_Salamanca_2013,Mahajan_2013,Socolovsky_2018}. At $z>1$ PSB galaxies are predominantly massive \citep{Wild_2016} and highly compact \citep{Almaini_2017}, while at lower redshift PSBs are predominantly low mass with morphologies consistent with low-mass spheroids \citep{Pawlik_2018,Maltby_2018}. This is consistent with a two phase quenching mechanism \citep{Wild_2016}, where at high redshift only high mass galaxies can quench quickly, \textcolor{mycolor}{while at low redshift rapidly quenching galaxies are found over a wide range in mass}. Hence PSBs potentially represent merger-induced quenching events that may highlight the role of gas-rich mergers and associated AGN feedback in quenching, but the importance of this pathway remains poorly quantified.


This paper aims to examine the relationship between mergers, starbursts and quenching within a large cosmological simulation, in order to better understand the various pathways to quenching and help to situate the above results within a hierarchical structure formation context.  We employ the new \simba\ cosmological hydrodynamic simulation~\citep{Dave_2019}, which utilises a novel black hole growth and AGN jet feedback model to yield an observationally concordant population of quenched galaxies.  Here we focus on understanding the connection between mergers, starbursts, and galaxy quenching, in order to quantify the contribution of the merger-driven quenching scenario to the overall population of quenched galaxies.  We quantify this by examining the quenching timescales of galaxies as a function of mass and redshift, and seek to identify a relationship between the quenching timescale and merger activity.  Along the way we also consider rejuvenated galaxies that were previously quenched but then returned to the star-forming main sequence.  Here we mainly focus on a theoretical investigation of mergers, starbursts, and quenching timescales; it is a rich and interesting study to examine whether the canonical observational signatures of these quantities are actually quantifying mergers accurately, but we defer this investigation to future work. 

This paper is organized as follows: in \S\ref{sec:sim_data} we briefly present the \simba\ simulation and describe our methods for tracking galaxies and quantifying mergers and quenching timescales. In \S\ref{sec:merger_analysis} we examine the enhancement in star-formation due to major mergers and their contribution to the cosmic stellar formation. In \S\ref{sec:quenching_analysis} the resulting quenching time-scale distribution is studied, and its connection to mergers and rejuvenations. Finally, in \S\ref{sec:conclusions} we present the conclusions extracted from the results.

\section{Simulation and analysis}\label{sec:sim_data}

\subsection{The \simba\ simulation}

We employ the \simba\ simulation suite for this analysis, which is described more fully in \citet{Dave_2019}.  \simba\ builds on the successes of the \mufasa\ suite of cosmological hydrodynamic simulations, which employed the \gizmo\ meshless finite mass hydrodynamics code \citep{Hopkins_2015} based on \gad~\citep{Springel_2005}, and includes state of the art sub-grid recipes such as $H_2$-based star formation following the sub-grid prescription of \cite{Krumholz_2009}, and chemical evolution tracking nine metals from supernovae and stellar evolution.  \simba, like \mufasa, employs a cosmology consistent with \cite{Planck_2016}, specifically: $\Omega_m= 0.3$, $\Omega_{\Lambda} = 0.7$, $\Omega_b = 0.048$ and $H_0= 68$ km s$^{-1}$ Mpc$^{-1}$. In this project, we employ the fiducial $100h^{-1}$ Mpc volume, with 1024$^3$ gas elements and 1024$^3$ dark matter particles. The simulation outputs 151 snapshots from $z=20\rightarrow 0$, with a snapshot spacing which is typically comparable to galaxy dynamical times ($\sim 100$~Myr at $z\sim 2$, increasing to $\sim 250$~Myr at $z=0$).

\simba's primary addition to \mufasa\ is a growth model of black holes and AGN accretion energy returned via bipolar outflows based on the observed dichotomy of feedback modes at high and low Eddington fractions~\citep{Heckman_2014}. Black holes grow through accretion of cold gas driven by gravitational torques \citep{Hopkins_2011, Angles-Alcazar_2013,Angles-Alcazar_2015} and accretion of hot gas following \cite{Bondi_1952}.  AGN outflows are implemented kinetically broadly similar to \cite{Angles-Alcazar_2017a}, using variable velocity and mass outflow rate to represent the transition from high mass-loaded radiativelly-driven winds to high-velocity jets -- this transition occurs at low Eddington ratios ($f_{Edd} < 0.02$) and high black holes mass ($M_{BH} > 10^{7.5} M_{\sun}$).  Other improvements include an on-the-fly dust evolution model \citep{Li_2019}, improved parameterisation of star formation feedback based on particle tracking in the Feedback in Realistic Environments simulations~\citep{Angles-Alcazar_2017b}, and improved cooling and self-shielding using the {\sc Grackle-3.1} library~\citep{Smith_2018}.

We identify and characterise the properties of galaxies using {\sc Caesar}, an extension to the {\sc yt} simulation analysis package~\footnote{caesar.readthedocs.io; yt-project.org.}.  Galaxies are identified via a 6-D friends-of-friends search, with a spatial linking length of 0.0056 times the mean interparticle separation, and a velocity linking length set to the local velocity dispersion.  {\sc Caesar} also calculates basic properties of the galaxy, such as $M_{*}$ and the total star formation rate (SFR), cross-matches it to a separate halo catalog (which we will not use in this analysis), and outputs a single hdf5 catalog file.  For this analysis, we only consider galaxies with a final ($z=0$) stellar mass of 
$M_*\geq 10^{9.5} M_{\sun}$, which corresponds to typically 175 star particle masses (or equivalently, initial gas element masses). This is larger than the nominal mass resolution corresponding to 32 star particles, but we employ a larger threshold in order to allow tracking galaxy growth back in time over a reasonable period.

\simba\ yields galaxies in good agreement with a wide range of galaxy~\citep{Dave_2019}, black hole~\citep{Thomas_2019} and dust ~\citep{Li_2019} properties. Relevant to this work, it produces a quenched galaxy population in good agreement with $z=0$ observations, both in terms of the sSFR vs. $M_*$ diagram, as well as the fraction of quenched galaxies as a function of mass.

\subsection{Identifying mergers}
\label{sec:merger_ident}

\begin{figure}
    \centering    \includegraphics[width=\columnwidth]{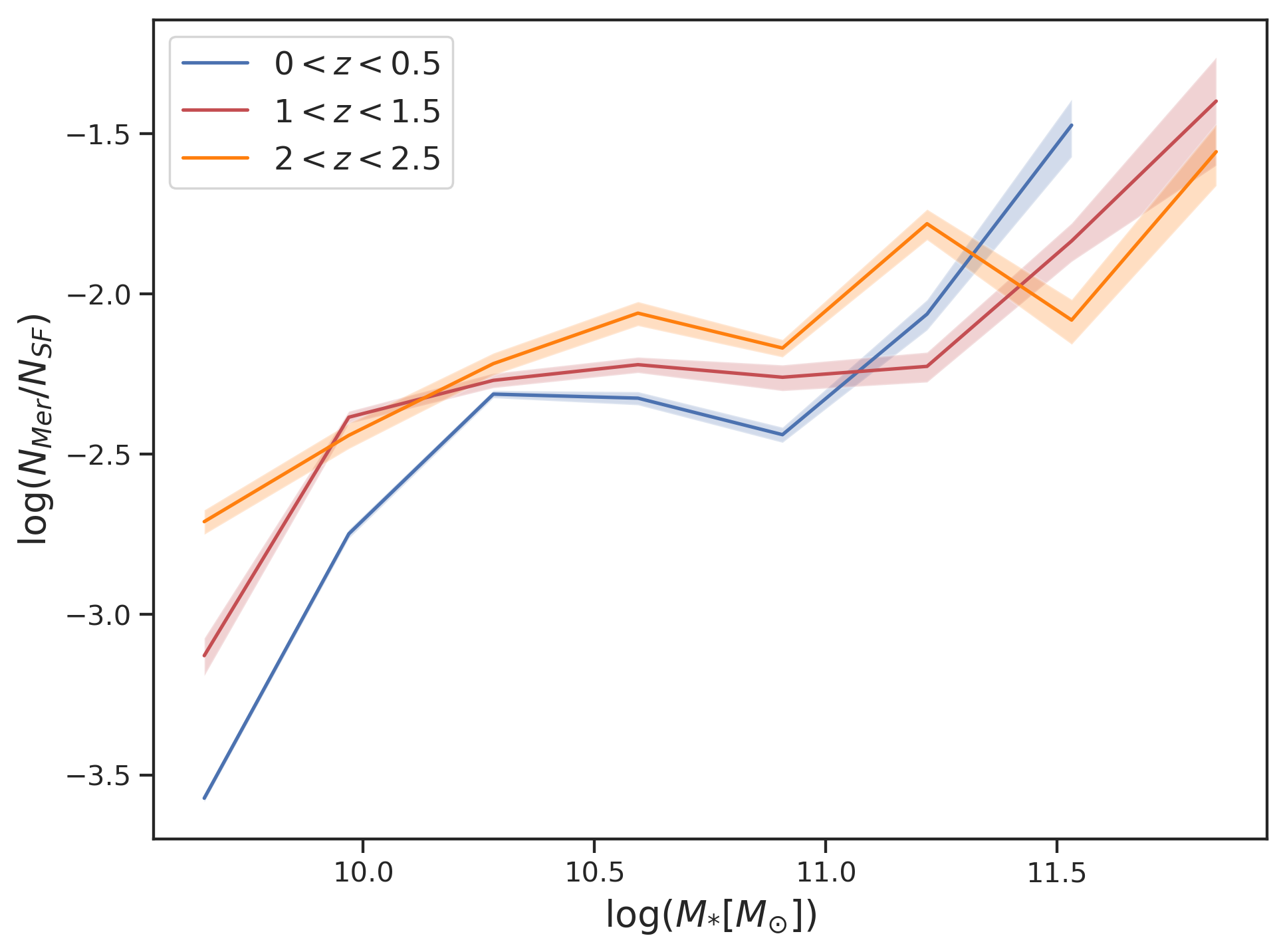}
    \caption{Fraction of major mergers in the star-forming population as a function of $M_*$, in three redshift bins: $0<z<0.5$ (blue), $1<z<1.5$ (red) and $2<z<2.5$ (orange).  The fraction of major mergers increases with $M_*$, and weakly with redshift.}
    \label{fig:merger_frac}
\end{figure}

To examine the growth of galaxies and identify merger events, we must track individual galaxies back in time through our snapshots.  To do so, we match up galaxies at the $z=0$ snapshot with galaxies in each previous snapshot by assigning the most massive progenitor to be the one containing the largest number of star particles in common.  With this, we can reconstruct the evolution of various galaxy properties for the main progenitor:  sSFR$\equiv$SFR$/{M_{*}}$, gas fraction $f_{H_2}\equiv M_{H_2}/M_*$, star formation efficiency SFE$\equiv$SFR/$M_{H_2}$, and type of galaxy. 

We define a major merger (which we will typically refer to as simply a merger) as a jump in the stellar mass of $\geq 20$\% relative to the previous snapshot, corresponding to a merger ratio $R\geq \textcolor{mycolor}{1:4}$.  Additionally, we require that the SFR at the previous snapshot be such that it would (if continued until the next snapshot) produce a mass increase that is less than 25\% of the total mass increase seen.  This latter criterion was tuned to avoid the early rapid growth phase of galaxies being incorrectly identified as mergers.  The specific numbers quoted in our results are mildly sensitive to this choice, but the overall trends and conclusions are not. To avoid ``fly-by" mergers or galaxy mis-identifications that occasionally happen in dense regions, we require that the previous three snapshots all satisfy the merger criterion relative to the post-merger snapshot.  We note that a rapid succession of minor mergers in between two snapshots could mimic a major merger; given our snapshot spacings we have no way to distinguish this.
Note that we ignore mergers between quenched galaxies (``dry" mergers), since we are interested in whether mergers are tied to galaxy quenching\textcolor{mycolor}{, i.e. we require that immediately after the mass jump, the remnant galaxy has a sSFR $> 0.2/\tH$.  We note that two quenched galaxies in principle could merge and result in a galaxy that enters our star-forming merger sample if it somehow acquires some cold gas to fuel star formation in the process, but this seems like an unlikely scenario}.  This sample of star-forming galaxy mergers is what we will use to investigate the connection between mergers and quenching. By applying this merger criterion, we find 507 mergers at $0<z<0.5$, 447 at $1<z<1.5$ and 189 at $2<z<2.5$. We choose these distinct redshift bins, which we will refer to as ``low", ``intermediate" and ``high", to identify evolutionary trends over cosmic time.  

The properties of the merging galaxy will be taken as the properties in the snapshot immediately after the mass jump.  Owing to the discreteness of snapshot outputs, this means that the actual merger could have happened anytime since the previous snapshot, so we are not typically catching the merger at its peak activity.   Nonetheless, once the merger begins, we expect any associated enhancement of star formation, etc., to continue for approximately a galaxy dynamical time.  Since our snapshot spacing is typically comparable to galaxy dynamical times ($\sim 100$~Myr at $z\sim 2$, increasing to $\sim 250$~Myr at $z=0$), we still expect to see some evidence of activity associated specifically with the merger.

Figure \ref{fig:merger_frac} shows the fraction of star-forming galaxies undergoing a merger as identified above, as a function of stellar mass, in our chosen redshift ranges.  We show $1\sigma$ jackknife uncertainties  as the shaded regions, computed as the cosmic variance over the merger fraction within 8 simulation sub-octants. The major merger fractions show a clear increase with stellar mass, at all epochs. This is expected given a reasonably tight relation between stellar mass and halo mass produced in simulations~\citep[e.g.][]{Agarwal_2018}, and the fact that the halo merger rate increase with halo mass~\citep[e.g.][]{Genel_2009} owing to hierarchical structure formation.  The overall numbers are fairly small, as $L^\star$-like galaxies at all redshifts considered have merger fractions of $\sim 1$\%.  These fractions are somewhat lower than \citet{Jogee_2009} found from examining morphologically disturbed galaxies at intermediate redshifts of a few percent, but this could owe to differences in merger selection.  We quantify merger rates in \S\ref{sec:quenching_analysis}. This sample of star-forming galaxy mergers is what we will use to investigate the connection between mergers and quenching.

\subsection{Quenching events and timescales}\label{subsec:quench_timescale}

\begin{figure}
    \centering
    \includegraphics[width=\columnwidth]{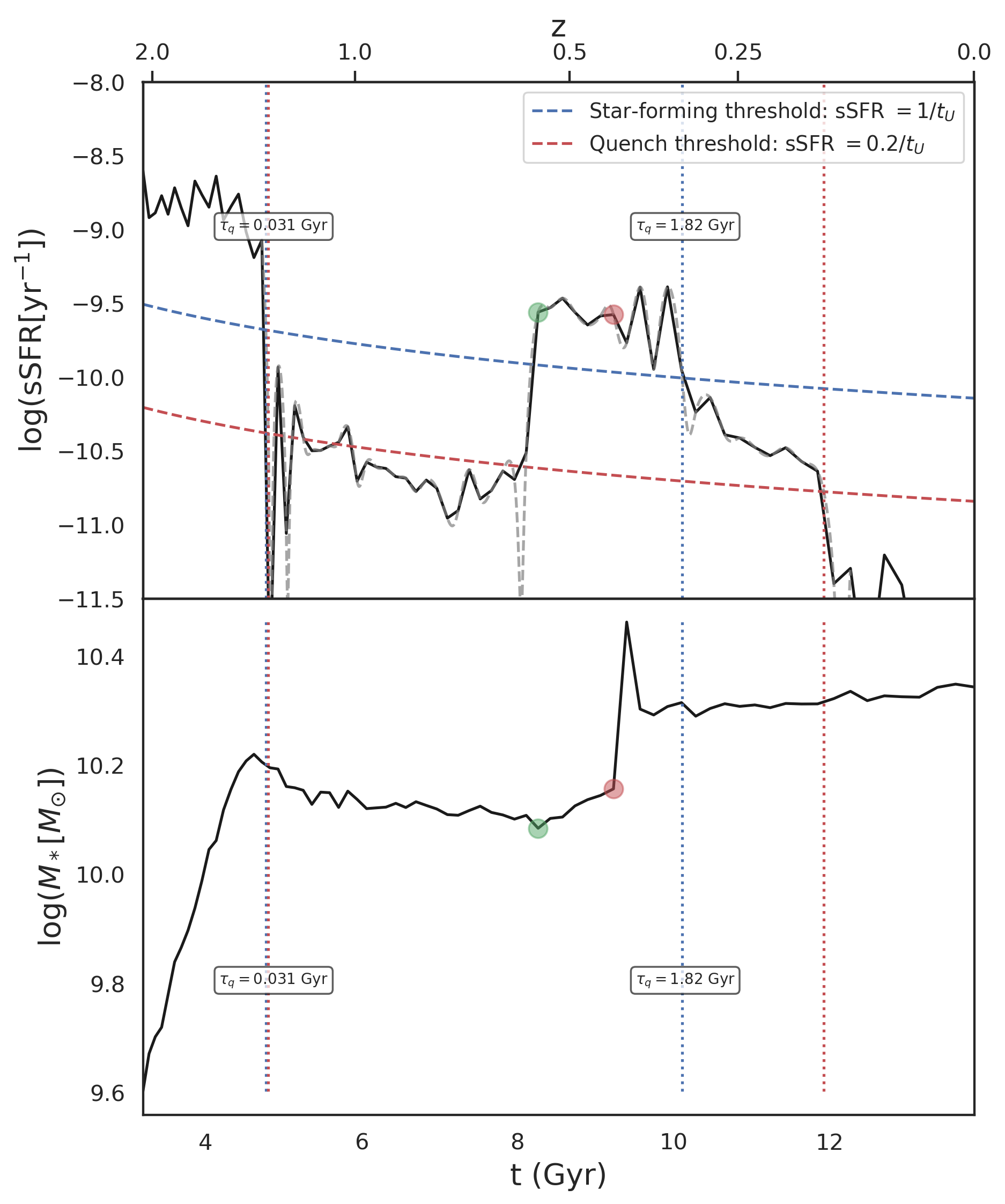}
    \caption{\textbf sSFR vs. time (top panel) and $M_*$ vs. time (bottom panel) for a chosen galaxy in the $100h^{-1}$ Mpc box. The dashed curves represent the star-forming (blue) and quenched (red) thresholds. Vertical lines indicate the start of the quenching process (blue dotted), and the end (red dotted). The time between these, the quenching time $\tau_q$, is shown in two little boxes close to the vertical lines at the start of quenching. A rejuvenation event is indicated by the green dot. Superimposed on the SFH, the dashed grey line shows the SFH of the galaxy as smoothed out by our spline interpolation technique, which is nearly indistinguishable. The starts and ends of the quenchings are given by the crossing of the interpolated curve, not the snapshot data. The galaxy also undergoes a major merger, indicated by the red dot.  This unusual galaxy shows two quenching events, one fast and one slow, with a rejuvenation in between, along with a merger.}
    \label{fig:quenching_example}
\end{figure}

To quantify quenching, we must first define it, then quantify the time required to quench. For simulated galaxies, quenching is most straightforwardly identified via thresholds in sSFR.  The value of the quenching threshold should vary with redshift, because galaxies show a dramatic evolution in their typical sSFR out to Cosmic Noon \citep[e.g.][]{Speagle_2014}.  Here, we employ the parametrisation proposed by \cite{Pacifici_2016}, in which a quenched galaxy at a given redshift $z$ is defined as one that has sSFR$(z)<\frac{0.2}{t_H(z)}$, where $t_H(z)$ is the age of the universe at redshift $z$. We analogously define a star-forming galaxy to be sSFR$(z)>\frac{1}{t_H(z)}$.  The range in between these is the green valley, which a galaxy must traverse in order to quench (see Figure \ref{fig:quenching_example}).  

A quenching event is considered to be when a galaxy drops from above the star-forming sSFR cut to below the quenched cut. Additionally, we require that the galaxy must stay below the star-forming cut (not the quenched cut) for at least another $0.2t_{H}(z_q)$, where $t_H(z_q)$ is the age of the Universe at the redshift of quenching $z_q$. We further define a galaxy to be rejuvenated if, after being quenched for a sufficient time, then has its sSFR return back over the star-forming cut. A galaxy is then eligible to quench again, so a given galaxy can have multiple quenching events, though as we will show this is uncommon.

The quenching time $\tau_q$ is defined as the time taken to go from above the star-forming cut to below the quenched cut. Owing to the discreteness of simulation snapshots, we compute this by first fitting a cubic B-spline \citep{Dierckx_1975} to the star formation history around the snapshot(s) where quenching occurs, and then determining the exact times when the sSFR crosses those two thresholds.

Figure \ref{fig:quenching_example} shows an example galaxy growth history chosen as an \textcolor{mycolor}{interesting} case where there are \textcolor{mycolor}{two} quenching events (shown by blue and red vertical dotted lines denoting the start and end, with $\tau_q$ as indicated), and thus \textcolor{mycolor}{one} rejuvenations in between (green dot). \textcolor{mycolor}{The red dot indicates a major merger. The blue and red dashed curves represent the star-forming and quenched galaxy thresholds.  The vertical blue and red lines show the beginning and ending of the quenching process, the difference being the quenching time $\tau_q$.  There are two quenching events, the first showing rapid quenching potentially associated with a flyby which causes some temporary sSFR fluctuations.  It is then quenched for a Gyr, and then undergoes a gas-rich minor merger ($\sim 1:10$) which boosts it back into the star-forming regime. It then undergoes a merger (red dot), after which it begins a slow decline towards quenching that takes nearly 2~Gyr.}  While such a rich history is uncommon, this does illustrate some potential connections between mergers, quenchings, and rejuvenations that we will attempt to quantify in this work.

Codes for merger identification and quenching time estimator are publicly available at \url{https://github.com/Currodri/SH_Project/tree/master}.

\section{The merger--starburst connection}\label{sec:merger_analysis}

\subsection{Global SFR in mergers vs. non-mergers}

We begin by examining the global connection between major mergers and enhanced star formation in the ensemble of \simba\ galaxies over cosmic time. It has long been observed that major mergers induce enhanced star formation~\citep[e.g.][]{Sanders_Mirabel_1996}, and simulations show that this owes to torques causing gas to lose angular momentum and flow inwards rapidly, causing a strong central burst of star formation~\citep[e.g.][]{Mihos_1996,Springel_2005b}. The amount of enhancement, however, depends strongly on the gas content of the merging galaxies and their detailed merger dynamics~\citep[e.g.][]{Cox_2007}.  While mergers are expected and observed to be more common at higher redshifts~\citep[e.g.][]{Lotz_2011}, observations suggest that the contribution of starbursts to the overall global SFR is modest even at Cosmic Noon~\citep[e.g.][]{Rodighiero_2011}.  In \simba, we can identify the population of merging galaxies at each epoch as described in \S\ref{sec:merger_ident}, and thereby examine the evolution of their typical properties over cosmic time.  By examining a representative population of galaxies relative to a subset that is merging at various epochs, we can thus quantify the contribution of mergers to the global star formation rate.

\begin{figure}
    \centering
    \includegraphics[width=\columnwidth]{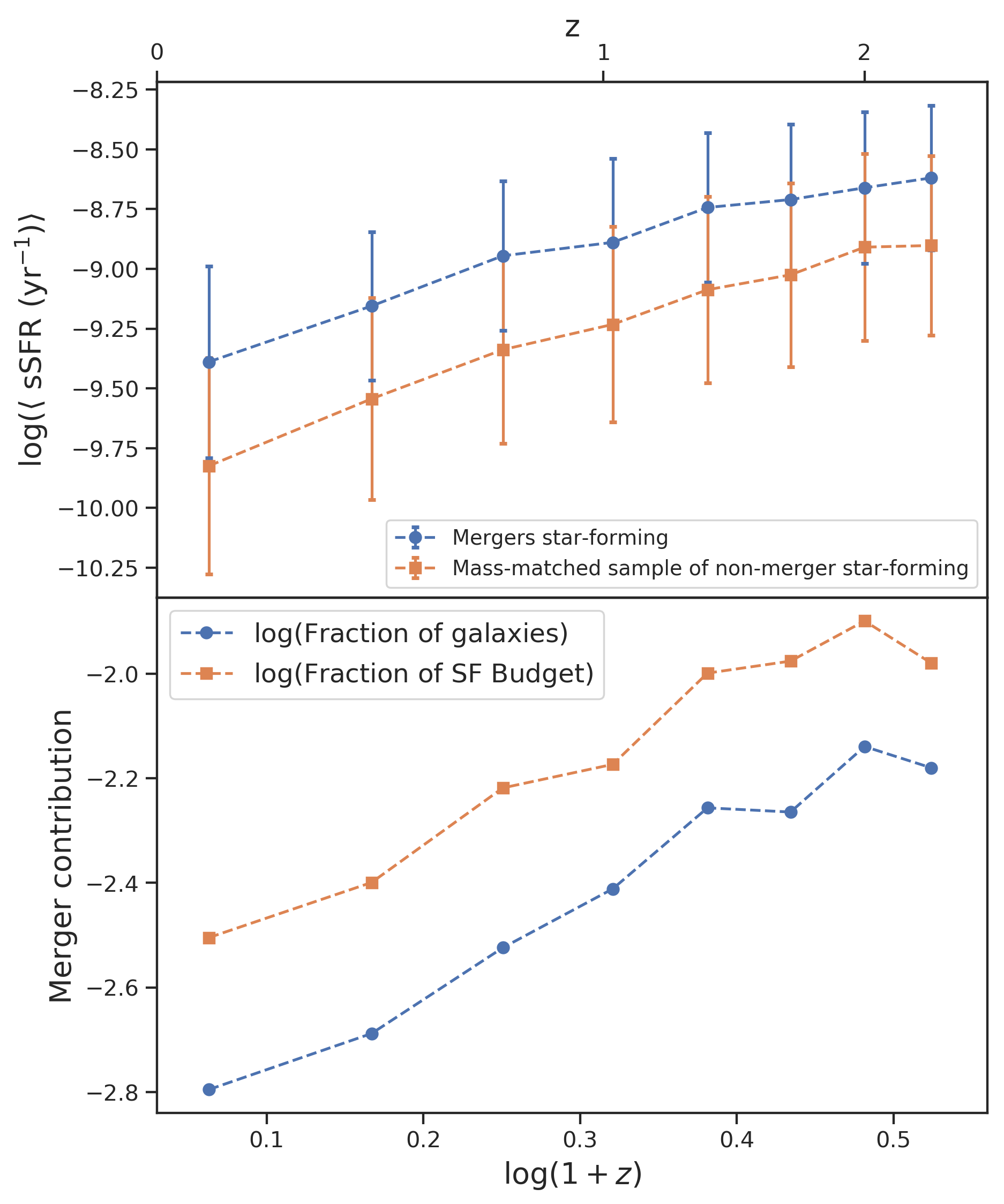}
    \caption{\textbf{Top:} Stellar mass weighted mean sSFR vs. redshift for star forming galaxies with $M_*>10^{9.5}M_{\sun}$. Blue points and line represent the running median of the merging star-forming galaxies, while the orange one provides the running median of mass-matched sample of non-merging main sequence galaxies. Errorbars are $1\sigma$ spread of galaxies around the mean in each bin.  \textbf{Bottom:} Fractional contribution of merging galaxies to the population of star-forming galaxies. The blue line represents the number fraction of mergers, while the orange line the fraction of the SF budget of star-forming galaxies provided by mergers. 
    }
    \label{fig:merg_sfr_evo}
\end{figure}

Figure~\ref{fig:merg_sfr_evo}, top panel, shows the stellar mass-weighted mean sSFR of all star-forming galaxies with $M_*>10^{9.5}M_\odot$ that have just undergone a merger as a function of redshift (blue), compared to a $M_*$-matched sample of star-forming galaxies that did not undergo a merger in that redshift bin (orange).  The uncertainties represent the $1\sigma$ range of sSFR values in each bin, showing a $\sim 0.3-0.4$~dex scatter in sSFR's as is typically observed~\citep{Kurczynski_2016}.  The stellar mass weighting means that this measure is most sensitive to galaxies near the knee of the stellar mass function, i.e. $M_*\sim 10^{10.5-11}M_\odot$.

The non-merger population displays the well-known dropping trend in sSFR with time, driven primarily by the dropping global mass accretion rate onto halos~\cite[e.g.][]{Dekel_2009,Dave_2012}.  The merger sample also shows a similar trend with redshift, but there is a clear enhancement in the sSFR amplitude relative to the mass-matched non-merger sample.  We note that the enhancement only appears in the post-merger snapshot; the pre-merger snapshot typically shows no or minimal enhancement.  The enhancement is fairly constant with redshift, and is $\sim\times 2-3$ in SFR.  This shows that mergers in \simba\ clearly drive enhanced SFRs.  

Figure~\ref{fig:merg_sfr_evo}, bottom panel, shows the fractional contribution of merging galaxies to the global star formation rate (orange) and the total galaxy number (blue), \textcolor{mycolor}{only considering the population of star-forming galaxies}.  Mergers contribute a percent or so to the global SFR budget at $z\sim 2$, dropping steadily to lower redshifts.  This is comparable to but somewhat lower than the 3\% contribution to the SF budget owing to the merger process from a study of morphologically disturbed galaxies at $z\sim 2$ in CANDELS data by \cite{Lofthouse_2017}.  \citet{Rodighiero_2011} found a somewhat higher contribution of $\sim 10\%$ from main sequence outliers (not necessarily mergers), but still a minor contribution to the global SFR density.  The number fraction is $\sim\times 2-3$ lower, commensurate to the typically $2-3\times$ enhancement in SFR in individual galaxies.  


\subsection{sSFR enhancement as a function of $M_*$}

\begin{figure}
    \centering
    \includegraphics[width=\columnwidth]{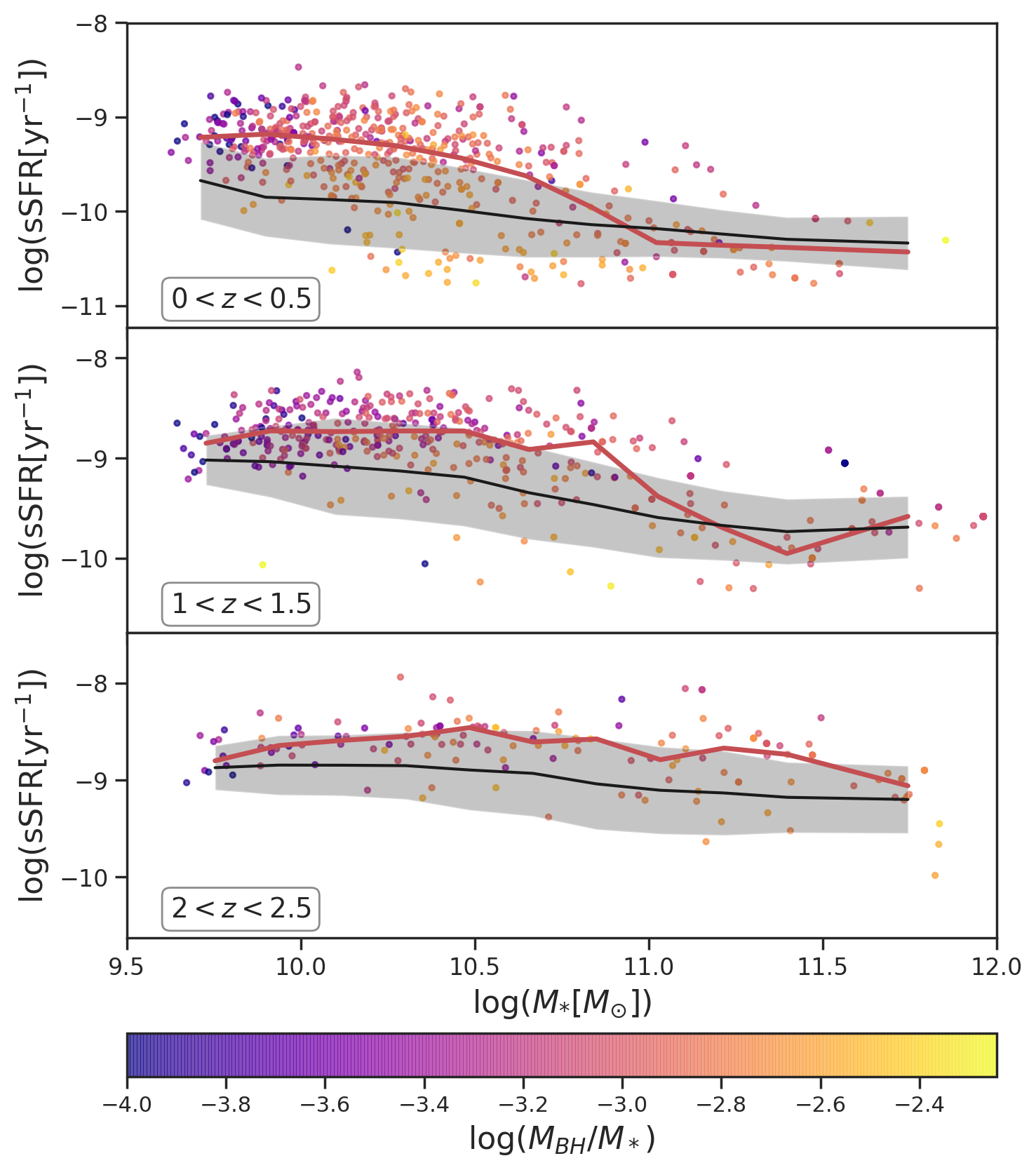}
    \caption{sSFR vs. $M_{*}$ in mergers identified via an increase of $\geq 20$\% in stellar mass relative to its progenitor at the previous snapshot (scatter points), within three redshift intervals (top to bottom panels). \textcolor{mycolor}{The merger points are colour-coded with $\log(M_{\text{BH}}/M_*)$ at the snapshot of the mass jump.} The median of the population of galaxies not undergoing mergers is shown as the black solid line, with the grey shaded region represents the $1\sigma$ deviation in the data.  The running median for the mergers is also shown in red.  Galaxies that have just undergone a merger generally show elevated star formation rates, but the ones with low sSFR tend to be ones with over-massive black holes.}
    \label{fig:merger_ssfr}
\end{figure}

The last section showed that mergers enhance star formation globally.  We now break this down as a function of stellar mass in our three chosen redshift intervals, in order to identify the particular galaxies where mergers are causing enhanced SFR, and thereby obtain insights into its physical drivers.

Figure \ref{fig:merger_ssfr} shows the galaxy main sequence, i.e. a plot of $M_*$ vs. sSFR, in three distinct redshift intervals: $0<z<0.5$ (top), $1<z<1.5$ (middle), and $2<z<2.5$ (bottom). The running median sSFR for the non-merger population of star-forming galaxies is represented by the black line, with the shaded region showing the $\pm1\sigma$ spread around the median. Here we only show star-forming galaxies, i.e. those with sSFR$>1/t_H(z)$.  Overlaid on this, we show as blue points individual galaxies that underwent a merger within that redshift interval, with the red line showing the running median sSFR of the mergers.

The overall main sequence in \simba\ behaves generally as observed:  It shifts downwards in sSFR and with an increasing (negative) slope to lower redshifts~\citep{Noeske_2007,Speagle_2014}.  We do not show an observational comparison here, but \citet{Dave_2019} showed that the $z=0$ main sequence agrees well with various recent determinations, and at $z\approx 2$ it is also in good agreement if one accounts for systematics in determinations of sSFR~\citep{Leja_2018}.  

Examining the merging galaxies, 
at each redshift it is clear that mergers cause enhanced star formation activity for all galaxies but those at the highest masses.  There appears to be a threshold mass below which enhancement occurs, which is $M_*\ga 10^{11.5}M_\odot$, dropping to $M_*\la 10^{11}M_\odot$ at low-$z$.  The typical enhancement is slightly greater than $+1\sigma$ relative to the median value, i.e.  $\sim\times 2-3$ relative to non-merging SFGs at a given $M_*$. We note that, since the actual merger occurs sometime before the snapshot where these properties are plotted, it may be that the peak enhancement in SFR is even larger than the enhancement seen.  Nonetheless, since the snapshot intervals are typically comparable to a disk dynamical time, it is still possible to see the impact of mergers on the SFR.  Our results are in qualitative agreement with CANDELS observations by \citet{Wuyts_2011}, showing that galaxies lying above the main sequence tend to show disturbed morphologies indicative of mergers. 

The lack of SFR enhancement at high $M_*$ presumably owes to the gas fractions in these galaxies being lower, and hence there is less fuel available to generate a starburst.  Furthermore, there are a few  galaxies that are actually within the quenched regime after the merger, particularly at lower redshifts.  This shows that occasionally mergers can indeed use up the gas and immediately leave a quenched remnant, as happens in isolated disk galaxy mergers~\citep[e.g.][]{Springel_2005b}.  \textcolor{mycolor}{We checked whether these high-massive galaxies stay quenched after the merger, and found that rejuvenations are very rare events, hence the vast majority of these objects do stay quenched until $z=0$.}

Overall, \simba\ shows a clear connection between major ($<1:4$) mergers, as identified by rapid mass growth, and enhanced star formation activity.  This enhancement is remarkably independent of stellar mass, once below a mass threshold that drops with time.  Next, we investigate more deeply the cause of the enhanced sSFR.

\subsection{What drives the enhanced SFR in mergers?}

\begin{figure}
    \centering
    \includegraphics[width=\columnwidth]{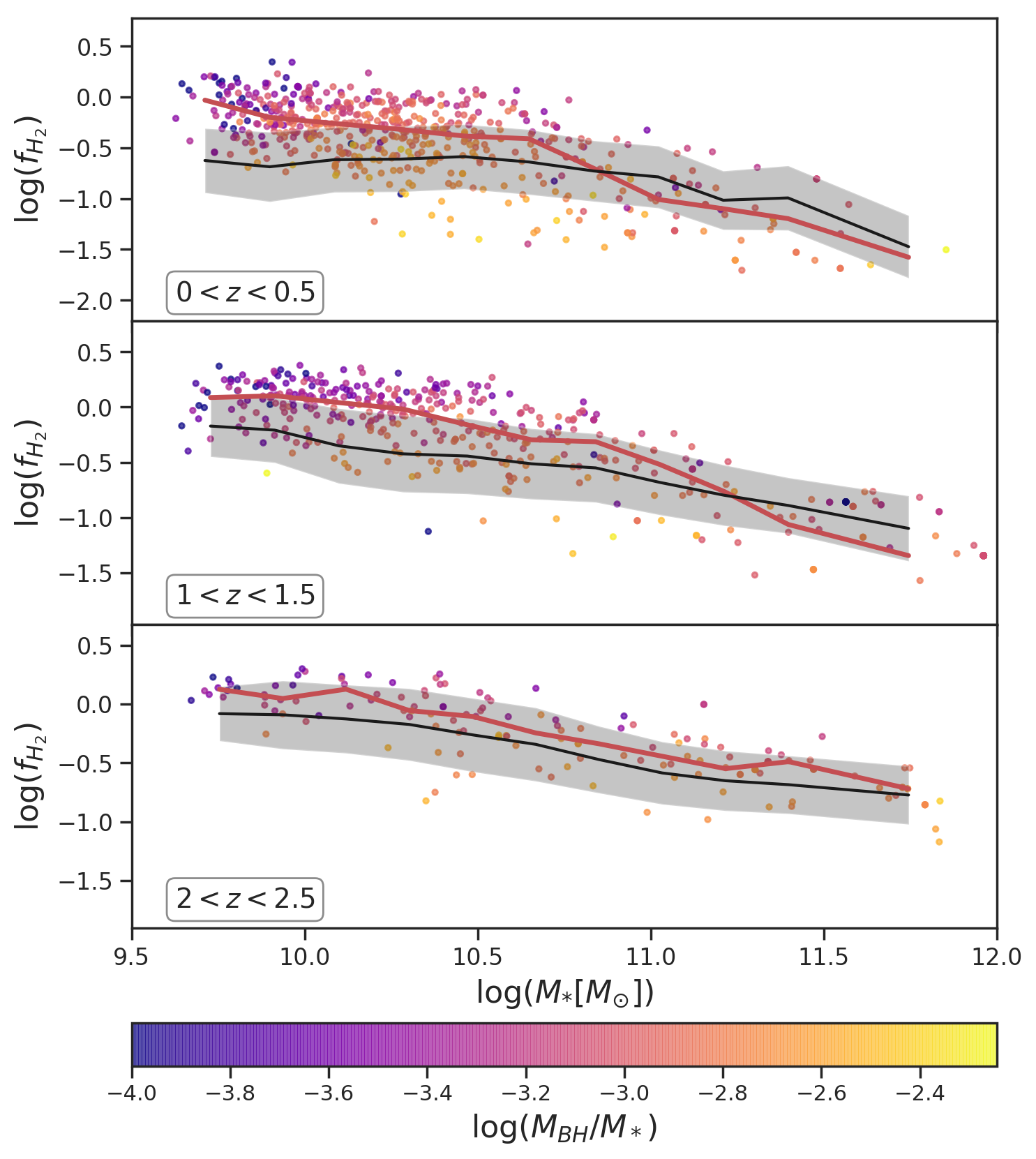}
    \caption{Logarithmic plot of the molecular gas fraction $f_{H_2}$ vs stellar mass $M_{*}$ in the mergers identified to be an increase of $\geq 20$\% in stellar mass. The data is organised in three redshift bins, showing the median of the population of galaxies not undergoing mergers (black solid line), with the grey shaded region representing the $1\sigma$ deviation in the data.  For merging galaxies, individual \textcolor{mycolor}{scatter points} are shown at the snapshot immediately after the merger\textcolor{mycolor}{, color-coded with $\log(M_{\text{BH}}/M_*)$ at the snapshot of the mass jump.}. The running median for the merger data is also shown as the red solid line.}
    \label{fig:merger_fgas}
\end{figure}

\begin{figure}
    \centering
    \includegraphics[width=\columnwidth]{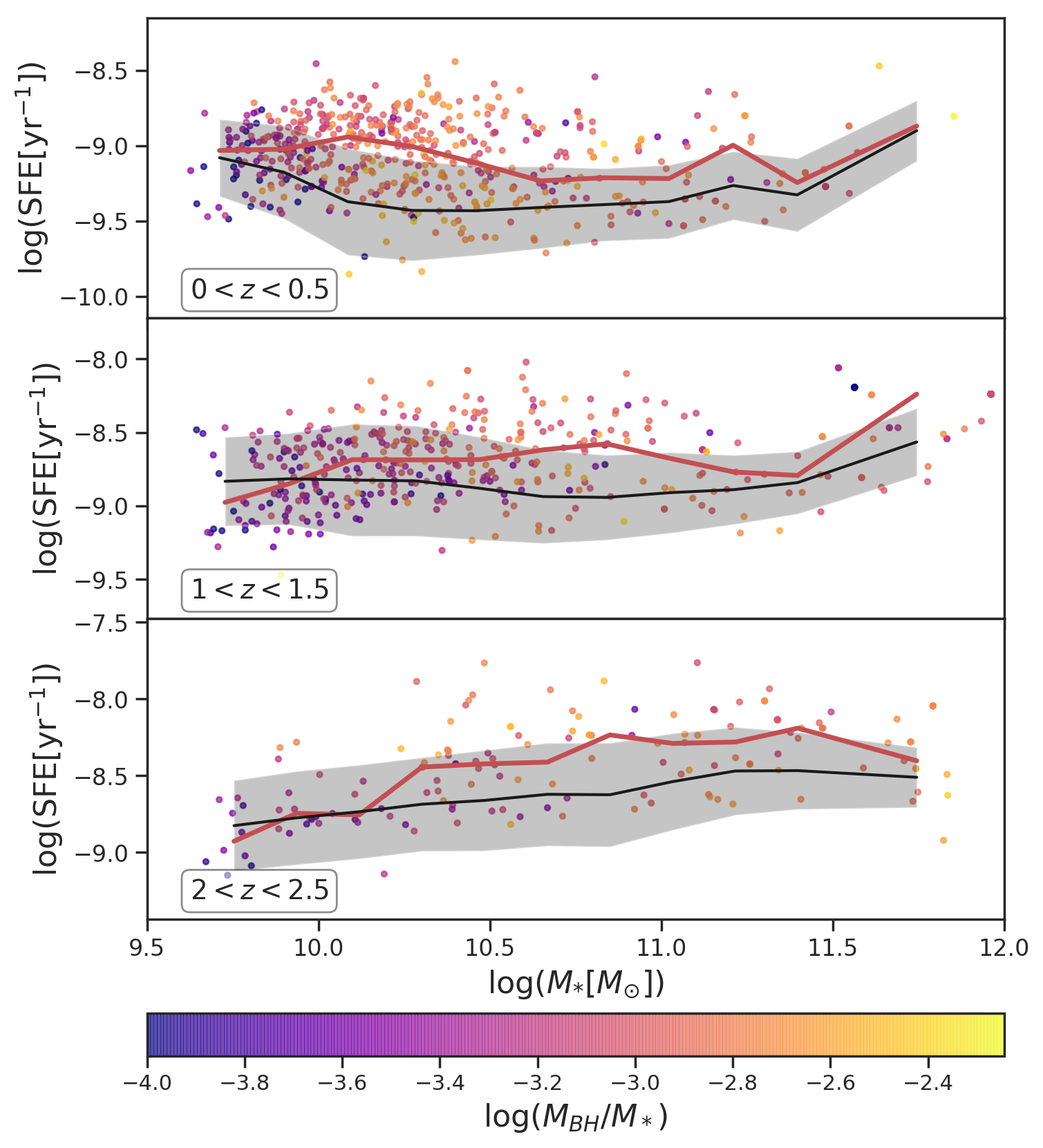}
    \caption{Logarithmic plot of the SFE vs stellar mass $M_{*}$ in the mergers identified to be an increase of $\geq 20$\% in stellar mass. The data is organised in three redshift bins, showing the median of the population of galaxies not undergoing mergers (black solid line), with the grey shaded region representing the $1\sigma$ deviation in the data. For merging galaxies, individual textcolor{mycolor}{scatter points} are shown at the snapshot immediately after the merger\textcolor{mycolor}{, color-coded with $\log(M_{\text{BH}}/M_*)$ at the snapshot of the mass jump.}.  The running median for the merger data is also shown as the red solid line.}
    \label{fig:merger_sfe}
\end{figure}

The sSFR can be written as sSFR$=\fmol\times$SFE=$\frac{M_{H2}}{M_*}\times\frac{\text{SFR}}{M_{H2}}$. Note that in \simba we manifestly require the star-formation to happen in molecular gas. Hence an increase in sSFR can owe to either an increase in $\fmol$ or an increase in SFE, or both.  Here, we investigate the cause for the enhanced sSFR in mergers by examining the enhancement in $\fmol$ and SFE separately for mergers with respect to the overall star forming population.

Figure \ref{fig:merger_fgas} shows the molecular gas fraction $\fmol$ in galaxies as a function of $M_*$ in three redshift intervals, analogous to Figure \ref{fig:merger_ssfr} for sSFR.  As before, the median of the overall SFG population is shown as the black line with $\pm1\sigma$ scatter about the mean shown in grey, while the merger galaxies are denoted by blue with a red line as the running median.

It is clear that mergers have an enhanced molecular gas fraction relative to the overall galaxy population.  The level of enhancement is approximately similar to that seen in the sSFR, showing that at first glance, much of the starburst nature of merging galaxies can be attributed to an increased fuel reservoir of molecular gas.  We will quantify this below.

The reason for the enhanced $\fmol$ could owe to several processes.  In \simba, as in \mufasa, molecular gas is tracked via a sub-grid prescription following \citet{Krumholz_2009}, which depends on density and metallicity.  Since the mass-metallicity relation is relatively tight~\citep{Dave_2019}, it is unlikely that enhanced metallicities could be responsible for increasing $\fmol$ at a given $M_*$.  Instead, it is more likely that the merger event causes gas to be funnelled towards the centre, drawing in \HI\ and potentially ionised gas from the outskirts that is then compressed to sufficient density to become molecular, as seen in galaxy merger simulations \citep[e.g.][]{Moreno_2019}.  In this way, mergers can drive enhanced molecular gas content which in turn spurs a starburst.  One would then expect the distribution of molecular gas in mergers to be more concentrated than in non-merging galaxies; we will examine this in future zoom simulations.  At low redshifts and high masses, a few post-merger galaxies are seen to have essentially zero gas content, and have correspondingly low sSFR; in these cases, mergers can immediately remove the star-forming gas and induce quenching, but this is rare.  In most cases, quenching (if it happens) occurs later on, as we will quantify below.

Figure \ref{fig:merger_sfe} shows the corresponding plot for star formation efficiency (SFE) as a function of $M_*$, in three redshift intervals.  Overall, one expects less enhancement in the SFE, given that much of the enhanced sSFR can be explained by the enhanced $\fmol$.  This is clearly seen to be the case, as SFE for merging galaxies is typically within the $1\sigma$ region of the non-merger sample. However, this isn't always negligible; particularly at lower redshifts at intermediate masses, the increase in SFE is comparable to that in $\fmol$.

The physical cause of an increase in SFE is likely driven by the density distribution of the molecular gas within galaxies. \simba's star formation recipe sets the SFR of a given particle to be $\propto \fmol\rho^{1.5}$~\citep{Schmidt_1959}, where $\rho$ is the gas density and $\fmol$ is, in this case, the molecular fraction of that individual particle.  In the case where the gas is dense and essentially fully molecular, the increased overall SFR within the galaxy must reflect an increase in the typical density of the ISM gas. This can occur owing to compression of the gas via the merging dynamics.  Hence it appears that for intermediate-mass gas-rich mergers at low redshifts, the distribution of the star-forming gas is altered to become more compressed, which drives an increased efficiency of conversion of molecular gas into stars.

\textcolor{mycolor}{To further explore the relationship between starbursts and mergers, the points have been colour-coded by ratio of the central black hole mass  to the stellar mass, $M_{BH}/M_*$. For the case of Figure \ref{fig:merger_ssfr}, it can be seen that, at a given stellar mass, galaxies with a bigger black hole tend to have a lower sSFR, even making some lie below the main sequence. This trend is also noticeable in Figure \ref{fig:merger_fgas}, where the equivalent can be said for $\fmol$, but not for the SFE in Figure \ref{fig:merger_sfe}. This suggest that, in merger in which the most massive galaxies holds a large black hole for its stellar mass, black holes can be related to a fast decline of the star-formation, primarily by affecting the molecular gas content after a merger. Further investigation into the effects of black holes on merger-driven starbursts will be considered in future work.}

It is worth commenting on numerical resolution.  Owing to \simba's force softening length of 0.5$\hkpc$ (comoving), it is not possible to fully resolve a dense central knot of star formation as seen in local starbursts such as the Antennae or Arp 220.  Thus one might find it surprising that \simba\ is nonetheless able to generate merger-driven starbursts. However, it is worth noting that the pioneering isolated merger simulations of \citet{Mihos_1996}, which produced quite strong bursts, had comparable spatial resolution, and a mass resolution that was only $5\times$ better than that of \simba.  While subsequent isolated merger simulations have substantially improved on this \citep[e.g.][]{Moreno_2019}, it is perhaps not surprising that mergers in \simba\ can still generate starbursts.  A consequence of this, however, is that it is possible that the peak strength of the starburst may be tempered owing to resolution effects.  We note that \citet{Moreno_2019} found a peak enhancement of up to $\sim\times 10$ in SFR during coalescence, so given that we are not usually catching the merger at peak enhancement given our snapshot intervals, a factor of $\sim\times 2-3$ seems plausible.  As a final caveat, we note that the enhancement seen in the SFE only in more massive galaxies may be partly a resolution effect, since the internal dynamics of gas can be better resolved in those galaxies in \simba; if we could resolve small galaxies at a similar level, perhaps we would also see a similar SFE enhancement there also.  Overall, while some of our detailed conclusions may be impacted by \simba's limited resolution, the broad trends are likely to be robust.

\subsection{Quantifying the merger-driven enhancement}

\begin{figure}
    \centering
    \includegraphics[width=\columnwidth]{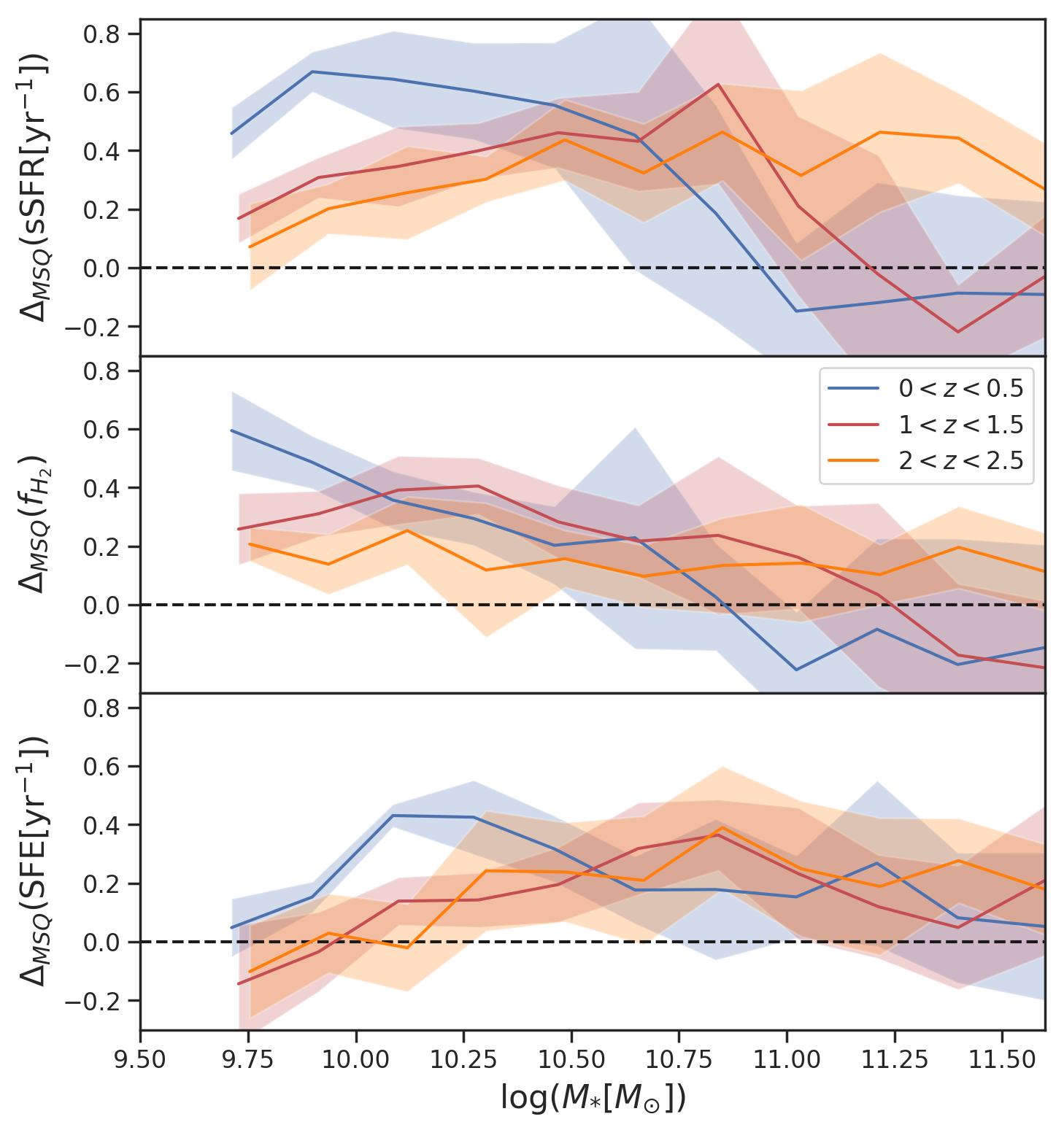}
    \caption{Logarithmic increase of sSFR, $f_{H2}$ and SFE of mergers with respect to the main sequence. The evolution with the three redshift bins is shown with different line colours: $0<z<0.5$ with blue, $1<z<1.5$ with red, and $2<z<2.5$ with orange. The horizontal black dashed line represents the main sequence, and the shaded regions represent the cosmic variance as obtained from dividing the simulation box in 8 equal octants. Results from the $100 h^{-1}$ Mpc box, with a sample of galaxies with stellar masses above $10^{9.5}M_{\sun}$ and classified as star-forming.}
    \label{fig:dist_gas}
\end{figure}

A comparison between the sample of mergers and non-merger SFGs shows that mergers have enhanced SFR, which is driven primarily by an increase in \textcolor{mycolor}{molecular} gas fraction but in certain cases can be comparably boosted by the SFE.  To more precisely quantify this, we consider the logarithmic deviations of the merging galaxies' median quantities from the \textcolor{mycolor}{star-forming} main sequence, $\dmsq$, as a function of $M_*$, in our three redshift intervals, for sSFR, $\fmol$, and SFE.  Given the definitions of our three quantities, we expect that $\dmsq$(sSFR)$\approx \dmsq(\fmol) + \dmsq$(SFE). Here, we investigate $\dmsq$ as a function of $M_*$, which quantifies the various contributions to the SFR enhancement.

Figure \ref{fig:dist_gas}, top panel, shows $\dmsq$(sSFR) as a function of stellar mass, within our high, intermediate, and low redshift intervals.  The shading shows the $1\sigma$ cosmic variance over the 8 sub-octants in the simulation volume.  As before, we see that the sSFR enhancement occurs up to a given $M_*$ whose value drops with redshift.  The peak enhancement generally shifts towards lower masses at later epochs, so that by $z=0$, galaxies with $M_*\approx 10^{10}M_\odot$ show a $\ga\times 3$ typical sSFR enhancement, whereas at higher redshifts this occurs closer to $M_*\approx 10^{11}M_\odot$.  For Milky Way-sized SFGs, mergers cause a $2-3\times$ enhancement at higher redshifts, but only a $50\%$ enhancement at low redshift.  This likely owes to the dropping gas fractions at a given $M_*$; at low redshifts, many high mass mergers are with very gas-poor galaxies, which do not provide additional fuel for a starburst.  At low redshifts, Milky Way-sized galaxies in \simba\ have gas fractions of $\fmol\ll 10\%$~\citep{Dave_2019}, as observed, whereas such low gas fractions are not typically the case for any SFGs at high redshifts since galaxies are overall more gas-rich (see Fig.~\ref{fig:merger_fgas}).

The middle panel of Figure \ref{fig:dist_gas} shows an analogous plot for the molecular gas fraction enhancement $\dmsq(\fmol)$.  The molecular gas enhancement is an inverse function of $M_*$ at all redshifts, even transitioning to a deficit of $\fmol$ for merging galaxies at $M_*\ga 10^{11}M_\odot$ at lower redshifts.  At the lowest masses, the molecular gas enhancement can essentially fully explain the sSFR enhancement, but not so at higher masses.  Hence the merger dynamics of drawing in gas into the molecular zone seems to be more effective in lower mass galaxies, likely because they have a larger reservoir of \HI\ gas~\citep{Dave_2019} that can be drawn in to molecular densities during a merger.  Meanwhile, the deficit at high masses shows the effect of consuming gas in a merger when the galaxies are initially relatively gas-poor, and typically living in hot halos where there is not a reservoir of cold gas to be drawn in.

The bottom panel of Figure~\ref{fig:dist_gas} shows the analogous plot for the SFE.  The enhancement in SFE occurs at intermediate masses for the high redshift bin, but shifts towards higher masses at lower redshifts.  If SFE enhancement is driven by the disruption of disk orbits to funnel gas into the centre as is canonically believed~\citep{Mihos_1996}, then such disruption requires the initial presence of an ordered disk.  Our results are then consistent with the idea that higher-mass star-forming galaxies are in ordered disks, while lower-mass ones are not, which itself is broadly consistent with observations of galaxy dynamics at both low and high redshifts~\citep[e.g.][]{Forster-Schreiber_2009,Kassin_2012}. We will investigate the detailed internal dynamics of mergers in future work using zoom simulations.

In order to quantify the statistical significance of these result in the SFE mass distribution, we perform a two-sample Kolmogorov-Smirnov test \citep{Peacock_1983} between the merger sample and the star-forming population, in each redshift bin and in 4 mass bins. The null hypothesis that \textcolor{mycolor}{the merging and non-merging star-forming galaxy populations are part of a common parent distribution} is rejected at $>95$\% confidence, i.e. with a p-value below 0.05, for all masses and redshifts except at low-redshift and high-$M_*$. 

In summary, \simba\ produces enhanced SFR in star-forming galaxy mergers by $\sim\times 3$ up to a stellar mass scale that evolves from $M_*>10^{11.5} M_{\sun}$ at $z=2$, to $M_*\la 10^{11}M_\odot$ at $z=0$.  The enhancement at the lowest masses is driven by an enhancement in the molecular gas fraction, likely owing to the dynamics of merging galaxies drawing in surrounding cold gas to increase the molecular reservoir.  At intermediate masses, the enhancement occurs in both $\fmol$ and SFE, suggesting that the gas within ordered disks is dynamically disrupted to yield a more concentrated starburst by driving a higher typical ISM density.  At the highest masses, the statistics are poor, but in general it appears that the SFE is strongly enhanced, which offsets the fact that $\fmol$ shows a deficit relative to the non-merger population. Galaxies rarely show immediate quenching after the merger.  In the next section we will examine whether such mergers are associated with galaxy quenching over a longer timescale.

\section{Connecting Mergers, Quenching, and Rejuvenations}\label{sec:quenching_analysis}

\subsection{Quenching time distributions}\label{subsec:quench_distribution}

Mergers enhance star formation, but do they also correlate with quenching?  And if so, do they induce rapid quenching, as canonically hypothesised~\citep[e.g.][]{Hopkins_2008}? To examine this, we will identify quenching events, and measure their quenching times.  Recall that we define a quenching event as a galaxy's sSFR going from above $t_H^{-1}$ to below $0.2t_H^{-1}$; the time spent in between those limits is defined as the quenching time $\tau_q$.  Here, we consider the 7947 galaxies that are found to be quenched at $z=0$ with $M_* \geq 10^{9.5}M_{\sun}$ (out of a total of 29716), and trace their evolution backwards in time to identify the redshift at which quenching occurred ($z_q$) and the time it took to quench $\tau_q$.  A total of 8281 quenching events were found, since a single galaxy can have multiple quenching events if it is rejuvenated in between; hence we find globally $\sim 1.04$ quenching events per $z=0$ quenched galaxy. 

\begin{figure}
    \centering
    \includegraphics[width=\columnwidth]{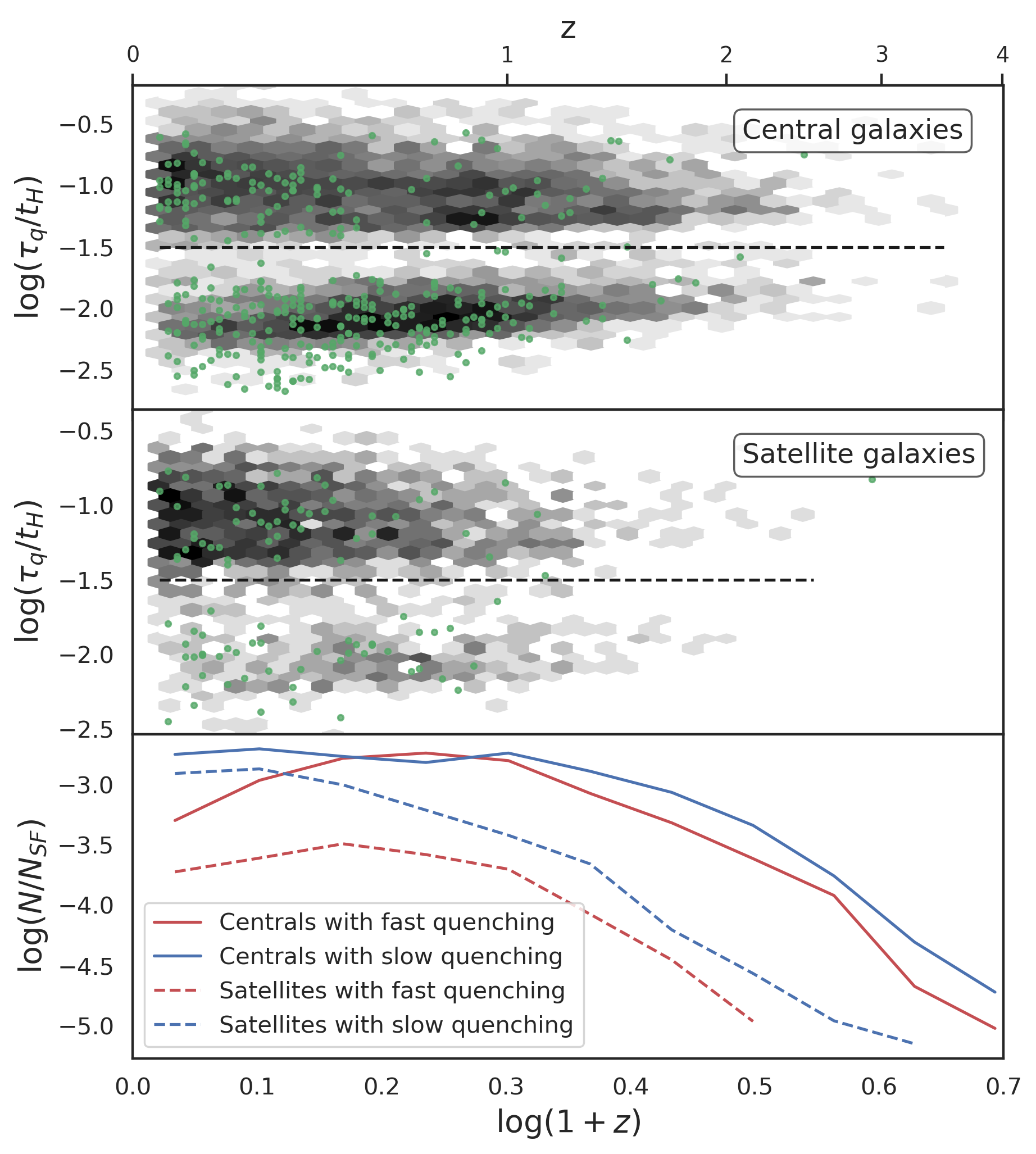}
    \caption{Final quenching times as a function of redshift, for $z=0$ quenched galaxies in \simba. \textbf{Top:} Hexbin plot of $\log\ t_{q}/t_H$ of central galaxies vs. $\log(1+z_q)$, with $z_q$ the redshift at which the quenching ends. Green circles represent galaxies that experienced a rejuvenation prior to this final quenching.  The horizontal black dashed line shows the demarcation between slow and fast quenching modes.
    \textbf{Middle:} Same as above but for satellite galaxies.
    \textbf{Bottom:} Fraction of star-forming galaxies that undergo quenching in each redshift bin, divided in centrals and satellites vs. $\log(1+z_q)$.
    }
    \label{fig:quenching_redshift}
\end{figure}

Figure \ref{fig:quenching_redshift} shows a hexbin plot of $\tau_q/t_H$ as a function of redshift from $z=4\to 0$, for central galaxies (top panel) and satellites (middle panel), as classified at the start of the quenching.  We only plot the final quenching redshifts and quenching times for these galaxies, ignoring the small number of quenchings that were subsequently followed by rejuvenations. Overlaid on this with individual green points are the subset of quenching events that underwent a rejuvenation prior to their final quenching; we will discuss rejuvenations later.

When scaled by $t_H$, the quenching time $\tau_q$ is starkly bimodal, with a division at $\tau_q/t_H=10^{-1.5}$; this division is shown as the horizontal dashed line for reference. There is no reason why our measurement of $\tau_q$ in \simba\ should disfavor identifying quenching times of $\tau_q\sim t_H/30$.  Thus it appears to be a physical bimodality predicted by \simba, and represents a key result of this paper.  We thus find a surprisingly clear distinction between fast and slow quenching events.  At face value, this is indicative of two distinct physical processes causing the quenching.

As an aside, we note that the earliest quenched galaxies in \simba\ appear at $z>3$.  It has long been a difficulty of hierarchical galaxy formation models to quench galaxies at sufficiently early epochs to match observations of the earliest quenched galaxies~\citep{Schreiber_2018}.  Indeed, the number density of quenched galaxies, when defined as in \citet{Schreiber_2018} as having sSFR$<0.15$~Gyr$^{-1}$, is $2.2\times 10^{-5}$~Mpc$^{-3}$ from $z=3-4$ in \simba, in very good agreement with their observed value of $2.0\pm 0.3\times 10^{-5}$~Mpc$^{-3}$. Hence while they point out that several other current galaxy formation models fail this test by an order of magnitude, this is not the case for \simba. 

The bottom panel of Figure~\ref{fig:quenching_redshift} shows the fraction of star-forming galaxies that undergo quenching in each redshift bin, for centrals (solid) and satellites (dashed).  For this, we also further subdivide this into fast quenching events (blue) and slow quenching events (red).

Galaxy quenching occurs increasingly frequently \textcolor{mycolor}{towards lower redshift}, flattening out since $z\sim 1$ at a rate of 0.1\% of star-forming galaxies with $M_*>10^{9.5}M_\odot$ undergoing quenching within each redshift bin.  The redshift trends are similar for fast and slow quenching -- specifically, \simba\ does not yield a trend that high-redshift quenching is preferentially in the fast mode, as some observations suggest~\citep{Wild_2016,Pacifici_2016}.  However, we note that this is the case when $\tau_q$ is scaled to $t_H$; since $t_H$ is obviously smaller at earlier epochs, $\tau_q$ is physically shorter at high redshifts. Our results suggest that scaling $\tau_q$ with $t_H$ provides a more consistent view of quenching times across all cosmic epochs.  This might be expected if quenching relates to the shutoff of accretion owing to e.g. halo gas heating~\citep{Gabor_2015}, which is a halo-related process and hence should occur on halo dynamical times that scale with $t_H$.

The majority of quenching events occur in central galaxies, but at late epochs, slow quenching satellites become comparable in frequency.  This likely occurs because satellites in massive halos can have additional quenching processes associated with starvation and stripping that are effective once large hot gaseous halos are in place.  In \simba, these appear to provide a boost in numbers to the slow quenching mode, while fast satellite quenching tracks the central galaxies more closely. In \citet{Rafieferantsoa_2019} it was shown using group zoom simulations that the satellite quenching time is long primarily owing to a delay period before the satellite recognises that it is living in a massive halo environment, whereas the final quenching event owing to e.g. gas stripping is rapid.  At face value, this seems contradictory to the direct measure of $\tau_q$ we do here, but it may be the case that such relatively massive groups are not the typical environments in which satellites are quenching.  We will examine the process of quenching satellites with respect to large-scale structure more generally in upcoming work (Kraljic et al., in prep.).

\begin{figure}
    \centering
    \includegraphics[width=\columnwidth]{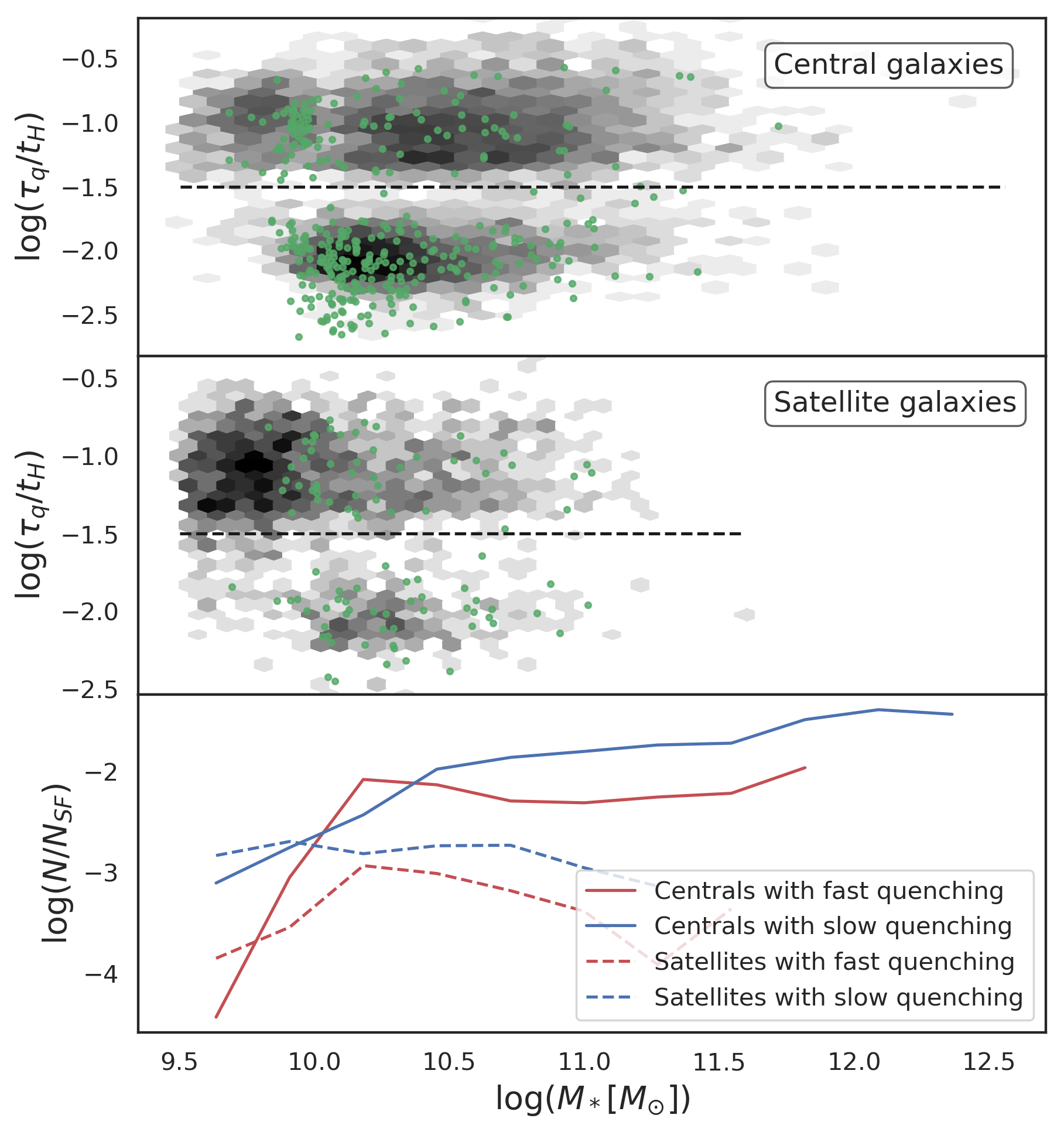}
    \caption{Final quenching times as in Figure~\ref{fig:quenching_redshift}, except now vs. stellar mass $M_*$.  Quenching times are strongly bimodal at $\tau_q/t_H\approx 0.03$.  For centrals, slow quenchings are somewhat more numerous overall, except around $M_*\sim 10^{10.3}M_\odot$.  Satellites are dominated by slow quenching at all masses.}
    \label{fig:quenching_mass}
\end{figure}

Figure~\ref{fig:quenching_mass} shows an analogous hexbin plot to Figure~\ref{fig:quenching_redshift}, except now as a function of $M_*$, taken over all redshift from $z=0\to 4$.  Again, we show the dividing line at $\tau_q/t_H=10^{-1.5}$, which breaks up the galaxy sample as before into fast and slow quenching events.  We again also show final quenching events that were preceded by rejuvenations as individual green points.  The bottom panel shows the fraction in fast and slow mode, for centrals and satellites, as a function of $M_*$.

For central galaxies (top panel), we see that not only is there a bimodality in $\tau_q/t_H$, but there is a clear shift of quenching from slow to fast mode right around $M_*\sim 1-3\times 10^{10}M_\odot$ (bottom panel).  Fast quenching thus clearly happen preferentially in this mass range.  Since mergers are not preferentially found in this mass range, this suggests that merging is not usually the origin of fast quenching in \simba.

This mass range does coincide with that where AGN jet quenching in \simba\ becomes effective~\citep[see e.g. Figure~3 of][]{Dave_2019}.  Hence it appears that fast quenching is more likely associated with the appearance of strong jet and X-ray feedback (which are approximately coincident in \simba).  The jet feedback is responsible for heating halo gas, and explicitly does not affect most star-forming gas in the ISM since it is ejected purely \textcolor{mycolor}{bipolarly}. Nonetheless, such jets could quench star formation on relatively short timescales if it causes a rapid cessation in accretion.  The X-ray feedback affects ISM gas by giving it an outward push in accord with the expected radiation pressure; this could hasten the removal of any remaining ISM gas.  \citet{Dave_2019} demonstrated that, overall, most of the quenching is enacted by the jet feedback, as the outwards push from X-ray feedback tends to be at relatively low velocities. Hence it is somewhat remarkable that such jet feedback, generally regarded as a preventative feedback channel and hence putatively a slow process, can actually act on relatively rapid timescales of $\sim 0.01t_H$.  Note that galaxies do not have to quench at the same time as their AGN jets turn on; it could be that the jet feedback heats surrounding gas, and then the rapid quenching occurs at some later time as the remaining gas is quickly consumed.

Galaxies that quench at higher masses tend to be preferentially in the slow mode.  This is consistent with the idea that they probably started quenching when the jets turned on, but did not complete quenching until they had grown in size over a longer period of time.  It is also the case that low-mass centrals quench almost exclusively in the slow mode.  It is unclear what is quenching such systems, but they could be the result of ``neighbourhood quenching"~\citep{Gabor_2015} where they happen to live near a massive halo and are impacted by the hot gaseous environment from a nearby large galaxy.

For satellites, the majority of quenching is in slow mode. There is still a bump in fast mode quenching at the same mass scale as for centrals, which in fact may owe to the vagaries of satellite vs. central definitions which depends on exactly how one identifies halos.  Generally, however, slow quenching appears to dominate in satellites.

Overall, \simba\ produces a bimodal quenching distribution, with slow quenching occuring at $\tau_q\sim 0.1t_H$ which is approximately on a halo dynamical time, and a fast quenching mode with $\tau_q\sim 0.01t_H$, with a dearth of intermediate quenching times.  This is suggestive of two distinct quenching mechanisms, which is reminiscent of the two-mode quenching scenario of \citet{Schawinski_2014} inferred from SDSS and ancillary data.  Their results suggest that massive late-type galaxies quench slowly, while lower-mass blue early-type galaxies quench more quickly. They speculate that the latter occurs owing to a rapid removal of gas such as might happen from a merger-induced starburst, which accompanies morphological transformation. The \simba\ results are qualitatively in agreement with the observations that more massive galaxies tend to quench more slowly, but suggest a peak in the fast quenching near the mass where AGN jet feedback turns on.  In \S\ref{sec:mergerquenching} we will examine whether this is tied to major mergers.

\subsection{Rejuvenation events}\label{subsec:rejuvenation_events}

Next, we consider the rejuvenation events.  These occur when a galaxy has been quenched, and subsequently returns back above the SFG threshold. This can occur because of a gas-rich major or minor merger, or potentially a buildup of a cold gas reservoir via cooling from hot halo gas.  In this section we examine properties of these rejuvenations, and how they relate to the final quenching events for those galaxies.

The green points in Figure~\ref{fig:quenching_redshift} and Figure~\ref{fig:quenching_mass} show the final quenching events that were preceded by a rejuvenation.  By eye, it is evident that these final quenchings are skewed towards fast mode.  While for the overall population the fraction of global fast mode quenchings is 37\%, for the rejuvenated galaxies it is 64\%.  In terms of mass and redshift the rejuvenations are not markedly different from the global population, although they are absent at the earliest epochs in the smallest galaxies mainly because rejuvenations require that galaxies first be quenched for at least $0.2t_H$.  However, the difference in terms of fast vs. slow mode is intriguing.

Rejuvenations appear to occur when a substantial reservoir of gas is added relatively quickly to the quenched galaxy. The example shown in Figure~\ref{fig:quenching_example} illustrates several such cases, showing a rapid increase in sSFR.  Hence rejuvenating galaxies do not evolve back slowly towards the main sequence.  This suggests that it may be associated with a particular event such as a merger, rather than a slow buildup of cooling halo gas; in that example, one of the rejuvenations is clearly associated with a minor merger, but the other one does not show any appreciable mass jump.  In the next section we show that there is no obvious correlation between rejuvenations and major mergers, hence the likely explanation is that these galaxies accreted a gas-rich satellite, flared up its star formation, and then quickly consumed its newfound gas owing to a combination of star formation and feedback.

The fraction of quenching events occurring in fast mode that are former rejuvenations is 10\%\textcolor{mycolor}{, with a median delay of $\sim 350$ Myr}.  While this is not a substantial population of the total fast quenching, if fast quenchings are associated with spectral signatures identified with post-staburst galaxies, then it could be that they represent a previously unrecognised class of PSBs that represent a rapid rejuvenation followed by a fast (re-)quenching.  This is different to the sort of burst traditionally envisioned for PSBs, in which it undergoes a gas-rich major merger that quickly transitions it from the SFG population to the quenched one \citep[e.g.][]{Peng_2010,Wild_2009,Wilkinson_2017,Poggianti_2009}.  In future work we will examine whether such rejuvenated galaxy fast quenchings do indeed have spectral signatures of post-starbursts, and whether they can be distinguished from more canonical post-starbursts.

In summary, quenching events in \simba\ split neatly into two categories of fast and slow, with a dividing line at $\tau_q\approx t_H/30$ independent of redshift.  The majority of quenching events are in slow mode, but fast mode is particularly prominent for galaxies with final masses of $M_*\approx 10^{10-10.5}M_\odot$, which is where \simba's AGN jet feedback starts to become effective.  Satellite quenching are rarer than central quenchings, but follow similar trends in redshift.  They are dominated by slow quenching, more strongly so towards later epochs.  Rejuvenations are rare, but preferentially arise in galaxies whose final quenching is in the fast mode.  These results suggest that AGN jet feedback plays a crucial role in initially quenching galaxies rapidly, but those that do not quench when the jets turn on end up quenching more slowly at higher masses.  The observational signatures of fast vs. slow quenching, and potentially those associated with rapid rejuvenations, is an interesting investigation that we leave for future work.  Next, we will more precisely quantify the evolution of these various processes over cosmic time.

\subsection{Merger, quenching, and rejuvenation rates}
\begin{figure}
    \centering
    \includegraphics[width=\columnwidth]{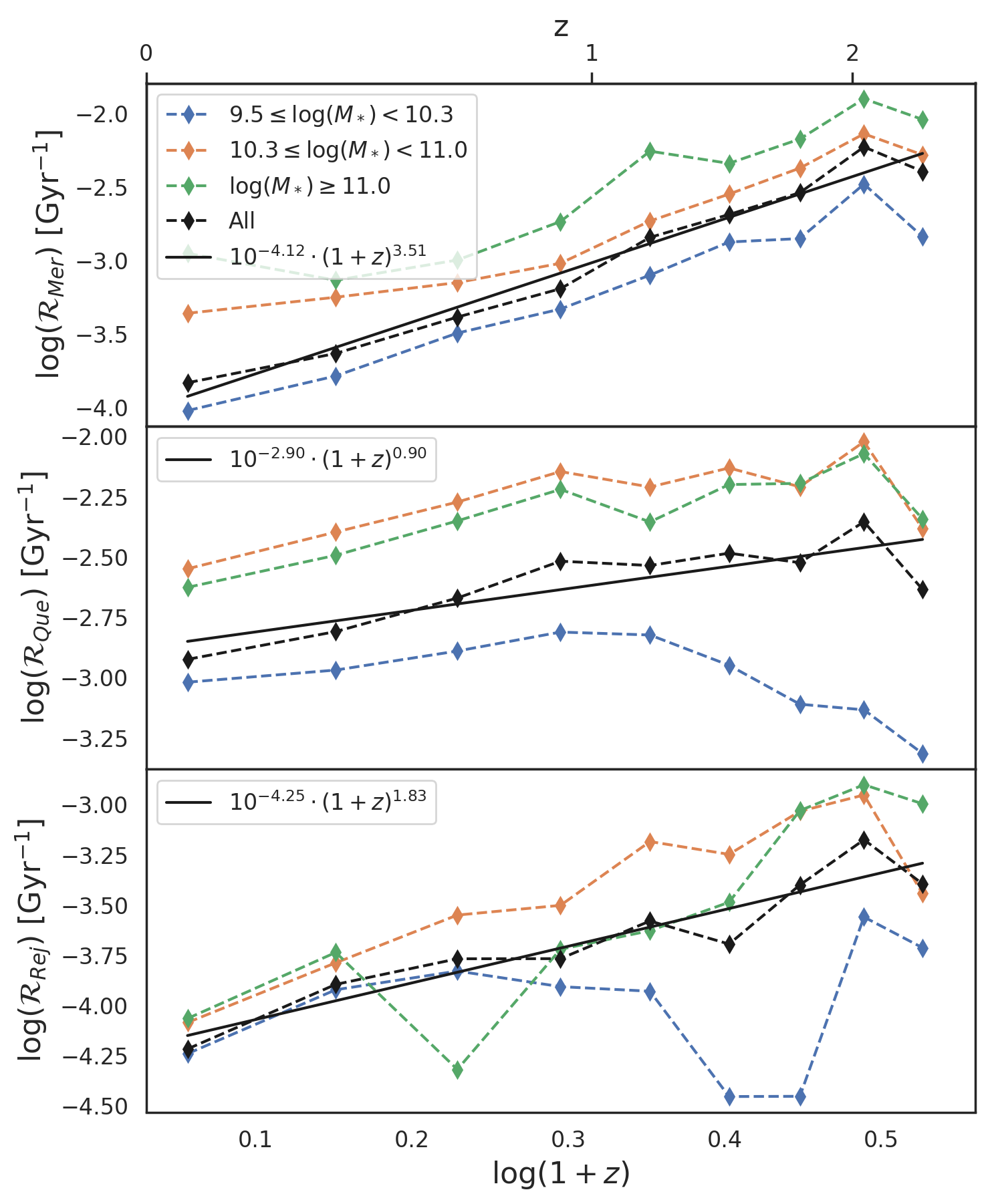}
    \caption{Evolution of the fractional rate of the three events studied in this work: mergers (top panel), quenchings (middle panel) and rejuvenations (bottom panel), vs. $\log(1+z)$. The black dashed line represents the total rates, while the coloured lines are the specific rates for each of the three mass bins studied: $9.5\leq \log(M_*) <10.3$ (blue dashed line), $10.3\leq \log(M_*) <11.0$ (orange dashed line), and $\log(M_*) \geq 11.0$ (green dashed line). To each of the total rates, we present the best power law fit with a solid black line.}
    \label{fig:fractional_rate}
\end{figure}

\begin{figure}
    \centering
    \includegraphics[width=\columnwidth]{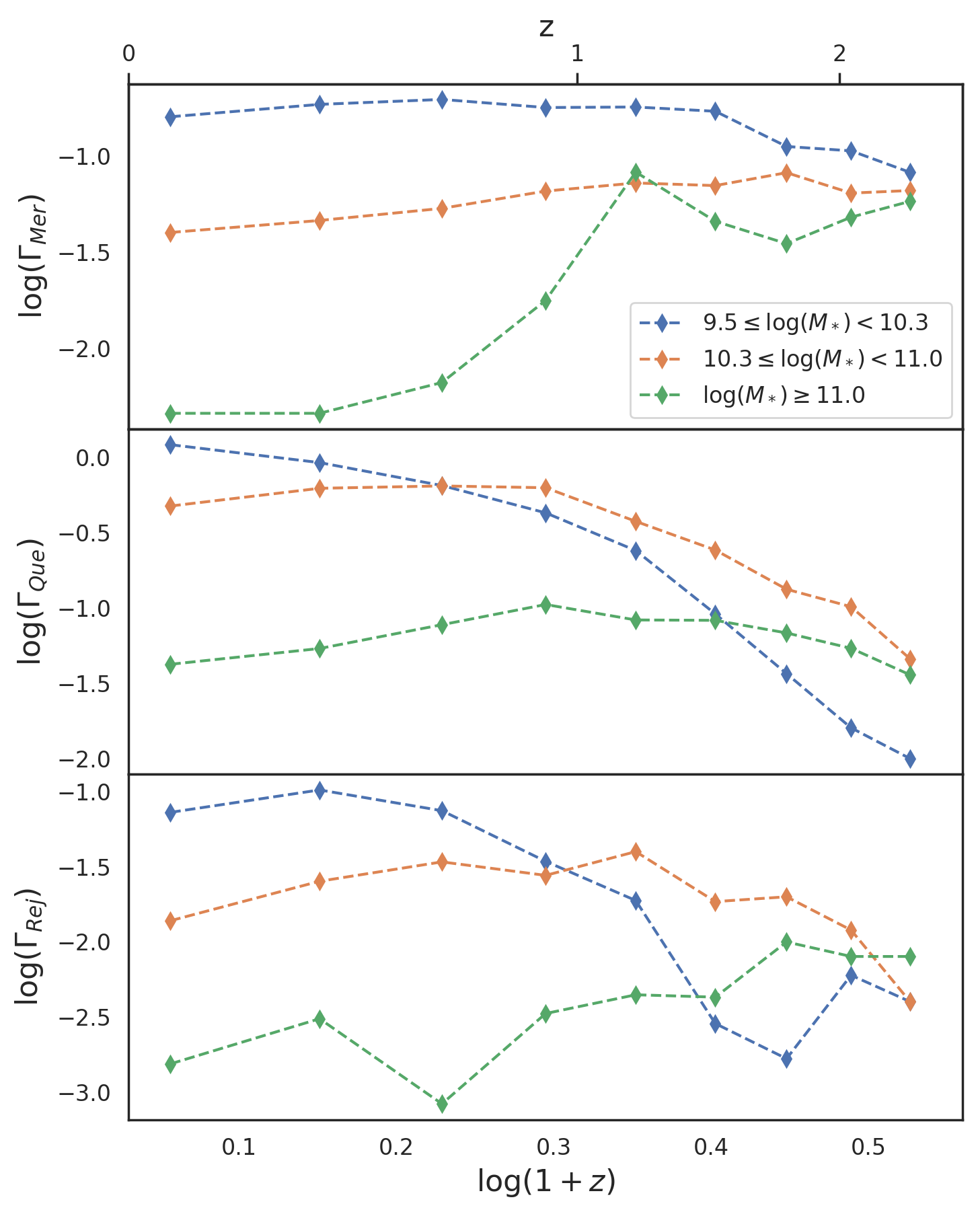}
    \caption{Evolution of the density rate (in cMpc$^{-3}$Gyr$^{-1}$) of the three events studied in this work: mergers (top panel), quenchings (middle panel) and rejuvenations (bottom panel), vs. $\log(1+z)$. The black dashed line represents the total rates, while the coloured lines are the specific rates for each of the three mass bins studied: $9.5\leq \log(M_*) <10.3$ (blue dashed line), $10.3\leq \log(M_*) <11.0$ (orange dashed line), and $\log(M_*) \geq 11.0$ (green dashed line).}
    \label{fig:density_rate}
\end{figure}

We seek to evolutionarily connect mergers to quenching and rejuvenations.  A broad view to this is provided by the evolution of their event rates over cosmic time.  A simple model in which each SFG merger results in a single quenched galaxy would predict the same amplitude and redshift trend for merger and quenching rates.  This is clearly too simple, as it is well established that if mergers are sufficiently gas rich, they will result in a star-forming descendant~\citep[e.g.][]{Robertson_2006,Hopkins_2008}.  Nonetheless, these rates provide an interesting guide to understanding the connection between these various events.

Figure~\ref{fig:fractional_rate} shows the fractional rate of events per Gyr for mergers (top panel), quenchings (middle), and rejuvenations (bottom), split into low, medium, and high $M_*$ bins (blue, orange, and green dashed lines, respectively).  Also shown is the total fractional rate for all galaxies combined (black dashed line), along with a best-fit power law (black solid line) with the fitting function indicated in the caption.  The fractional rate is the fraction of galaxies undergoing one of the events studied (merger, quenching or rejuvenation) per Gyr compared to the overall SFG population at that epoch\textcolor{mycolor}{, which we will refer to just as the ``rate".} Within each mass bin, the fractional rates are computed with respect to the total SFG population in that mass bin.

The merger rate for star-forming galaxies (top panel) shows a rapid decline with redshift, well-fit as $\propto (1+z)^{3.5}$.  This is comparable to the observations of \citet{Lotz_2011} who determined a scaling of $\sim(1+z)^{3}$ from a close pairs analysis. The steep dependence of the merger rate on redshift is a natural consequence of the hierarchical assembly of halos~\citep{Genel_2009}. There is also a clear mass dependence such that more massive galaxies merge more frequently, as already seen in Figure~\ref{fig:merger_frac}, which is exacerbated at late epochs.

The quenching rate (middle panel) shows a much weaker redshift dependence than the merger rate, $\propto (1+z)^{0.9}$.  This is expected because mergers are more frequent at high redshift, while quenching is more frequent at low redshift.  Nonetheless, it is interesting that when normalized per Gyr, the quenching rate actually still increas\textcolor{mycolor}{es} towards high redshifts.  Obviously this cannot continue indefinitely, as eventually there are no quenching events at $z\ga 4$ in \simba; nonetheless, over the redshift range probed here, quenchings are more frequent in time at high vs. low redshift.

There is also a strong mass dependence of quenching, as expected.  In particular, quenchings is much more common in galaxies with $M_*>10^{10.3}M_\odot$ than in lower-mass systems. Recall from Figure~\ref{fig:quenching_redshift} that most low-mass quenching events are in satellites, which are both less frequent overall and are preferentially at lower redshifts; this explains the small values in the low-mass (blue) trend and the lack of low-mass quenching at higher redshifts.  Meanwhile, the quenching rate is roughly independent of mass for $M_*>10^{10.3}M_\odot$, and the quenching fraction is even slightly higher for intermediate-mass galaxies.  These intermediate-mass quenchings are preferentially in fast mode, while the high-mass galaxy quenching are dominated by slow mode, but overall both evolve similarly with redshift.

It is instructive to compare the evolutionary rates of mergers versus quenchings, i.e. the top two panels of Figure~\ref{fig:fractional_rate}.  Not only is the redshift dependence markedly different, but comparing the amplitudes it is notable that the overall rate of quenchings exceeds that of mergers for $z\la 1$, and in higher-mass systems for $z\la 1.5$.  Hence there are simply not enough major merger events to explain the number of quenchings over the majority of cosmic time.  This is strong evidence that in \simba, major merging is not directly tied to quenching for most quenched galaxies.

The bottom panel of Figure~\ref{fig:fractional_rate} shows the rejuvenation rate.  In amplitude, this is considerably smaller than either the merger or quenching rate, as expected.  The evolutionary trend is somewhat closer to mergers, $\propto (1+z)^{1.8}$, which at face value may indicate a stronger connection between rejuvenations and mergers than quenchings and mergers.  There is a strong mass trend at high redshifts such that there are essentially no rejuvenations at low mass\textcolor{mycolor}{, since quenchings are rare in low-mass high-redshift galaxies}, but at low redshifts the rates in the three mass bins become comparable.

Figure~\ref{fig:density_rate} provides a different view on these same quantities.  In this plot, rather than normalising to the number of SFGs, we normalise to the total volume.  Since \simba\ has a constant volume in comoving Mpc$^3$, effectively this is a plot showing the evolution in number of these three types of events.

Overall, there is a much larger number of SFGs in the lowest mass bin (e.g. 7518 at $z=0$) vs. the intermediate and high mass bins (569 and 19 at $z=0$), respectively.  This is particularly true at lower redshifts when the majority of massive galaxies are quenched.  Hence the low-mass (blue) curves now dominate by number particularly at late epochs.

The relative trends with mass are, nonetheless, informative.  Looking at the volumetric merger rates, there are roughly a constant number of mergers per Gyr per comoving Mpc$^{3}$ over the redshift range probed here.  The high mass number density of mergers drops primarily because there is a decline in the fraction of that population that is star-forming. Hence viewed in terms of overall counts (rather than fraction of the population) gives a different view of which galaxies are dominating merger counts.

The volumetric quenching rate (middle panel) shows clear differences in evolution as a function of $M_*$. It is roughly constant per Gyr for the most massive galaxies, and increases mildly for intermediate mass galaxies before flattening at $z\la 1$.  Low-mass quenched galaxies, meanwhile, which are mostly satellites, strongly prefer quenching at later epochs.  This highlights the different quenching processes associated with centrals, mostly related to AGN feedback, versus satellites, where environmental processes are thought to dominate quenching.

Rejuvenations (bottom panel) are generally much rarer, with intermediate mass systems typically showing one rejuvenation per Gyr per 100~cMpc$^{-3}$.  Massive galaxies almost never rejuvenate except at the earliest epochs, which is expected since gas-rich mergers may provide enough fuel to boost the sSFR of a moderate-mass galaxy, but would need to be quite gas-rich in order to sufficiently boost the sSFR of a massive galaxy.  Low-mass galaxies also rejuvenate relatively frequently at $z\la 1$.  If they are mostly satellites, it is not entirely clear where such galaxies obtain substantial gas in order to do this, since it seems unlikely that they would be able to accrete it from surrounding halo gas.  Instead, it is more likely related instead to gas-rich mergers potentially in the outskirts of larger halos or prior to becoming a satellite.  It would be interesting to investigate the spatial distribution of such systems, which we leave for future work.

Rejuvenations are a process that has not been studied in large number observationally. Signatures of these have been found in radio galaxies \citep[e.g][]{Saikia_2007,Konar_2012}, as well as in elliptical galaxies \cite[e.g][]{Zezas_2003, Mancini_2019}. \cite{Mancini_2019} found that massive green valley galaxies have very old bulges but young disks, suggesting that rejuvenation owes to fresh accretion of new gas, not late-time bulge formation.  Our results are broadly consistent with this picture, because \simba\ rejuvenations do not come from major mergers and so would not be expected to strongly grow young bulges.

Overall, the fraction of galaxies undergoing mergers evolves steeply with redshift, while those undergoing quenching evolve much more slowly (but are still more frequent per Gyr than at low redshifts).  Major mergers are insufficiently frequent to explain moderate-mass or massive quenched galaxies at $z\la 1.5$.  Galaxies seem to quench at similar rates at all masses above $\sim 10^{10.3}M_\odot$. Tracking these events in number or number density (rather than fraction) shows distinctly different evolution for low-mass quenched galaxies (which are predominantly satellites) versus higher mass ones, reflecting different quenching processes.  Rejuvenations are generally rare, and occur in intermediate mass galaxies independent of redshift, and in low mass galaxies much more frequently at low redshift.  While these global trends suggest that mergers and quenching are not intimately tied in \simba, it is possible to examine this more precisely by relating merger and quenching events for individual galaxies, as we do next.

\subsection{Do quenchings follow mergers?}\label{sec:mergerquenching}

We can identify quenching events and merger events for individual quenched galaxies.  It is thus possible to examine on an individual basis whether quenching events follow merger events in \simba.  While the global trends described in the previous section suggest no overall connection between mergers and quenching, there could be a subset of quenching events that come relatively soon after a merger, indicating some physical connection, along with another population of quenched galaxies whose quenching does not correlate with merger activity. We thus seek to identify these populations, if they exist, and thereby quantify how often mergers enact quenching.  Moreover, since we can identify quenching as fast or slow based on $\tau_q/t_H$, we can see if mergers preferentially enact fast quenching. 

\begin{figure}
    \centering
    \includegraphics[width=\columnwidth]{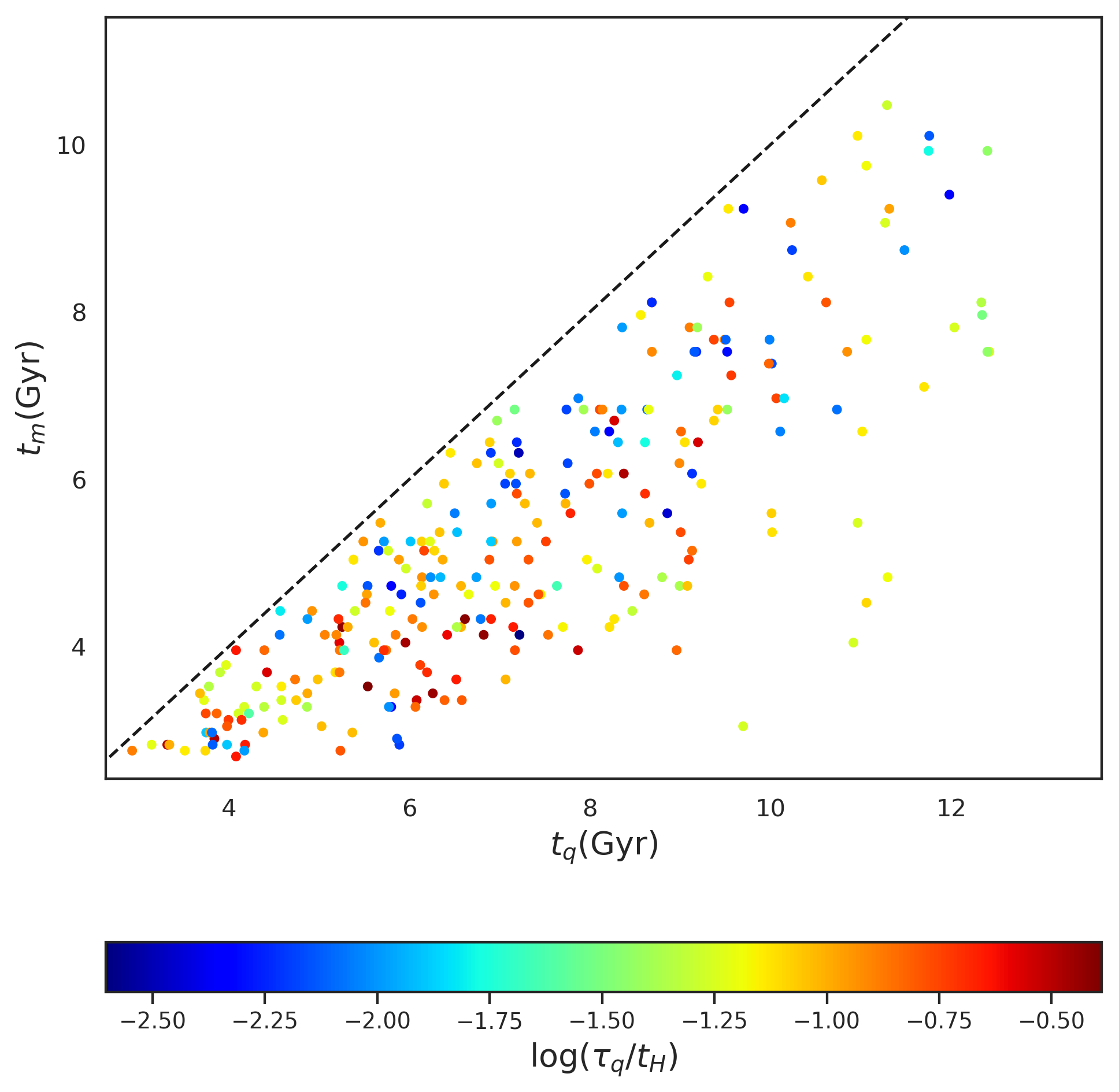}
    \caption{The time at which a merger happens in a star-forming galaxy $t_m$ vs. the time at which a quenching event starts \textcolor{mycolor}{$t_q$}, in Gyr. The scatter points are colour-coded with the logarithm of the quenching time $\tau_q/t_H$. The dashed black line represents the situation in which the merger occurs at the same snapshot as the quenching ($t_q=t_m$), i.e. quenching is instantaneous.}
    \label{fig:merger_time_quenching}
\end{figure}

\begin{figure}
    \centering
    \includegraphics[width=\columnwidth]{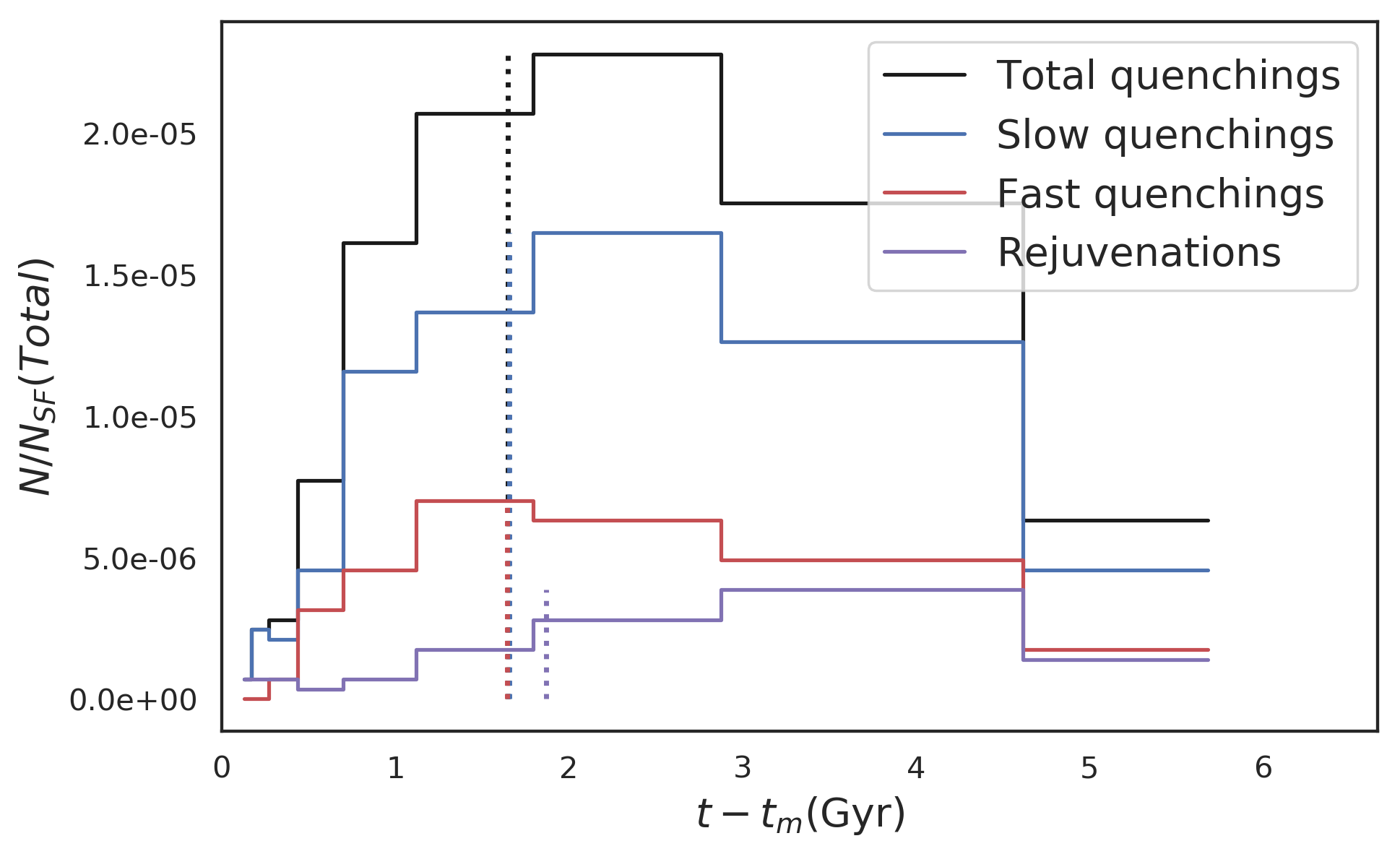}
    \caption{Histogram with the distribution of delay times $t-t_m$ between the times of quenching (blue line for slow quenching, red for fast and black for the total set of quenching events) and rejuvenation (magenta line) with respect to mergers. The number of events in each bin is scaled with the number of star-forming galaxies in all snapshots $N_{SF}(\text{Total})$. Vertical dotted lines represent the median of each distribution.  Quenching and rejuvenations both have typical delay times that are significantly longer than 1~Gyr, showing that they are unlikely to be related to the merger.}
    \label{fig:delay_time_histogram}
\end{figure}

Figure~\ref{fig:merger_time_quenching} shows a scatter plot of the time of merging $t_m$, in Gyr since the Big Bang, vs. the time of quenching $t_q$.  Note that $t_q$ is different to $\tau_q$; the latter is the duration of quenching, whereas the former is the cosmic time at which \textcolor{mycolor}{the onset of }quenching occurs.  Here we only consider galaxies that undergo a major merger, and quench at some later epoch; this represents 275 galaxies out of the total of 7947 quenched galaxies at $z=0$ -- the vast majority do not have a major merger while in the star-forming phase.  Finally, we colour-code the points by $\tau_q/t_H$, in order to examine any potential trends of mergers inducing fast quenching.  The one-to-one relation is shown as the dashed black line; galaxies close to this relation will have quenched very shortly after the merger.

The biggest takeaway from Figure~\ref{fig:merger_time_quenching} is that there are no clear trends, of any sort. There is not an obvious pile-up of quenching events close to the one-to-one line, as would be expected if the mergers were preferentially inducing quenching.  There is also no discernible trend with $\tau_q/t_H$, such that quenching events closer to mergers are preferentially undergoing fast quenching as might be expected\textcolor{mycolor}{; while the bimodality of quenching timescales is clearly evident from the colour-coding, there is no clear correlation of this with distance from the one-to-one line.}  This lack of correlation further solidifies the view that, in \simba, major mergers do not obviously drive fast quenching, or indeed any quenching, even in the small subset of quenched galaxies that have undergone mergers.

A more quantitative view is provided by a histogram of the delay time between the merger event and the quenching event.  This is shown as the black curve in Figure~\ref{fig:delay_time_histogram}. There is a median delay time of $\approx 1.5$~Gyr indicated by the vertical black dotted line. This is long compared to the duration over which a merger is expected to quench the star formation, suggesting that there is no physical connection.  Notably, there is no hint of a separate population of galaxies that quench very close to the time of the merger, as this would be expected to give a peak at small delay times.  

The histogram is also subdivided into fast (red) and slow (blue) quenching.  These show essentially identical distributions to the overall quenching.  Specifically, there is no tendency for fast quenching to occur closer to the time of the merger. 

Figure~\ref{fig:delay_time_histogram} also shows a histogram of the rejuvenation delay time relative to the merger (for the cases where rejuvenations occur after a merger).  The typical time delay is $\sim 2$~Gyr, again suggesting no physical connection.  Also, there is no clear evidence of a peak at small delay times, though there are a couple of galaxies overall that rejuvenate at the time of the merger.

In summary, \textcolor{mycolor}{major} mergers do not appear to be directly related to either quenching or rejuvenation events in \simba. It is not entirely clear why the quenching delay time peaks at 1.5~Gyr, but this could occur purely owing to a random distribution of quenching times after a given merger, up to the age of the universe.  There is no tendency for fast quenching events to be preferentially associated with mergers.  Hence in \simba, mergers appear to generate starbursts, but are unrelated to galaxy quenching.


\section{Conclusions}\label{sec:conclusions}

We have investigated the connection between galaxy mergers ($>1:4$), starbursts, quenching, and rejuvenations in the \simba\ $100\hmpc$ cosmological galaxy formation simulation. We define mergers based on a stellar mass jump that significantly exceeds that expected from continuous star formation, and we define quenching based on crossing from the star-forming population with sSFR$>t_H^{-1}$ to the quenched one with sSFR$<0.2t_H^{-1}$, with the quenching time defined as the time taken to cross the green valley in between. Rejuvenations are quenched galaxies that then return above the star-forming sSFR limit. In brief, our main result is that major mergers trigger elevated star formation activity, and we identify a \textcolor{mycolor}{clear} bimodality distribution of quenching times, but there is no obvious connection between merging and either fast or slow quenching.

Our main conclusions are detailed as follows:
\begin{itemize}
    \item Galaxies identified as just having undergone a merger show specific star formation rates $\sim\times 2-3$ higher than non-merging galaxies at all cosmic epochs from $z\sim 2.5\to 0$.  Nonetheless, they make up just a couple percent of the total cosmic SFR at high redshifts, dropping to even lower contributions by $z=0$.
    \item The enhancement in SFR connected with mergers occurs in all galaxies at high redshifts, but is restricted to galaxies with $M_*\la 10^{11}M_\odot$ at low redshifts, presumably because massive low-$z$ galaxies lack sufficient gas to trigger a burst.
    \item The enhancement is driven primarily by an elevated star-forming (molecular) gas fraction at lower masses, with an increasing contribution from an elevated star formation efficiency at high masses.  In massive galaxies at low redshifts, the post-merger gas fractions are actually lower than for non-merging galaxies, suggesting gas consumption.
    \item Quenching times in \simba\ are distinctly bimodal, with fast and slow mode quenching neatly divided at $\tau_q/t_H\approx 1/30$ at all cosmic epochs.  This is true for both central and satellite galaxies.
    \item For central galaxies, slow mode quenching is overall more common than fast mode, but in the mass range $M_*\sim 10^{10}-10^{10.5}M_\odot$, fast mode is more prevalent.  This correlates with the mass range where AGN jet and X-ray feedback becomes effective in \simba, providing a circumstantial connection between fast quenching and AGN feedback.
    \item Satellites have a lower quenching fraction, and are even more dominated by slow mode, particularly at late epochs.  This likely owes to environmental processes such as stripping and starvation being more important in these systems.
    \item \simba\ produces its first quenched galaxy at $z\approx 4.5$, and from $z=3-4$ the quenched galaxy number density is $\approx2\times 10^{-5}$~Mpc$^{-3}$, in very good agreement with observations~\citep{Schreiber_2018}.
    \item The fraction of star-forming galaxies undergoing merging events evolves strongly with redshift, as $(1+z)^{3.5}$.  The fraction of quenching events evolves much more slowly, at $\propto (1+z)^{0.9}$.  Mergers become too infrequent to explain all quenched massive galaxies at $z\la 1.5$.  Rejuvenations are rare, but subsequently tend to quench in the fast mode; their event fraction evolves as $\propto (1+z)^{1.8}$.
    \item By number, the counts of mergers are dominated by small galaxies, but for quenching they are typically dominated by intermediate-mass galaxies except at late epochs.  The volumetric quenching rate of galaxies is fairly constant of redshift, except for small systems (mostly satellites) which quench late.  The rejuvenation rate of intermediate mass galaxies is fairly constant, but low mass galaxies increase rapidly in time and dominate in number at $z\la 1$.
    \item Examining the time delay between the time of major mergers and that of final quenching shows no obvious correlation.  The typical time between merging and final quenching is $\ga 1$~Gyr, suggesting they are not physically connected.  There is also no evidence that quenchings occurring close (temporally) to mergers are preferentially fast mode events.
\end{itemize}

These results point towards some basic physical scenarios.  First, major mergers have a significant effect on the star-formation history of a main-sequence galaxy by driving strong variations in the gas content and distribution.  Second, such events are not responsible for quenching galaxies in \simba, and instead quenching appears to be more related to the onset of AGN jet feedback.  The detailed dynamics that gives rise to fast quenching from \simba's AGN jet heating along with X-ray radiation pressure is not immediately clear, though it may be that the onset of jets starves the galaxy, which then later undergoes a final burst that uses up its remaining gas.  Rejuvenations seem to be related to minor mergers with gas-rich satellites.  The life history of quenched galaxies is thus complex, and impacted by a range of physical processes related to hierarchical growth, star formation, and feedback.

While \simba\ is state of the art, there are a number of caveats related to numerics.  The first is numerical resolution, which may be insufficient to fully resolve central starbursts within major mergers.  The likely effect is to reduce the peak bursts strength, which may underestimate the importance of merger-driven starbursts.  Another is our limited time resolution, which does not enable us to track the merging and quenching process over sub-dynamical timescales; again, the sense of the effect is to reduce the statistical impact of the merger-driven burst, and compromise the measurements of very short quenching times.  For these reasons we might expect that the trends identified regarding starbursts in \simba\ may be a lower limit to the true effect.  It is also possible that, owing to the inability to resolve bursts and the heuristic nature of sub-grid star formation and feedback, \simba\ is somehow suppressing the effects of quenching associated with internal dynamics, such as compactification processes seen in zoom simulations~\citep{Tacchella_2016}.  Though this is unlikely to enact permanent quenching without an additional preventive feedback process~\citep{Gabor_2012}, it is possible that such processes may occasionally result in a closer connection between mergers and fast quenching.  Unfortunately, obtaining a statistical sample of massive galaxies to study quenching while retaining high resolution to resolve burst dynamics is very computationally challenging.  

In the future, we would like to better understand physically how AGN jets enact slow vs. fast quenching, and why it has the dependence on stellar mass that it does.  At present, it is unclear why there are these two distinct quenching modes, when there is only the single physical process of AGN feedback that appears to be ultimately responsible for quenching.  Also, we would like to examine the different processes associated with central and satellite galaxy quenching, particularly the role of large-scale structure environment in the latter.  Finally, it would be interesting to look for particular observational signatures of fast vs. slow quenching such as post starbursts features, and relate these back to physical processes in the simulation.  \simba\ provides a unique and state of the art platform to examine these and other physical processes associated with merger-driven starbursts and galaxy quenching.

\section*{Acknowledgements}

The authors acknowledge helpful discussions with Weiguang Cui, Katarina Kraljic, and Frazer Pearce. The authors thank Robert Thompson for developing {\sc Caesar}, and the {\sc yt} team for development and support of {\sc yt}.
RD acknowledges support from the Wolfson Research Merit Award program of the U.K. Royal Society.
DAA acknowledges support by the Flatiron Institute, which is supported by the Simons Foundation. DN and is supported in part by NSF Award AST-1715206 and HST Theory Award 15043.0001.  
This work used the DiRAC@Durham facility managed by the Institute for Computational Cosmology on behalf of the STFC DiRAC HPC Facility. The equipment was funded by BEIS capital funding via STFC capital grants ST/P002293/1, ST/R002371/1 and ST/S002502/1, Durham University and STFC operations grant ST/R000832/1. DiRAC is part of the National e-Infrastructure.




\bibliographystyle{mnras}
\bibliography{mergers} 

\begin{thebibliography}{}
\makeatletter
\relax
\def\mn@urlcharsother{\let\do\@makeother \do\$\do\&\do\#\do\^\do\_\do\%\do\~}
\def\mn@doi{\begingroup\mn@urlcharsother \@ifnextchar [ {\mn@doi@}
  {\mn@doi@[]}}
\def\mn@doi@[#1]#2{\def\@tempa{#1}\ifx\@tempa\@empty \href
  {http://dx.doi.org/#2} {doi:#2}\else \href {http://dx.doi.org/#2} {#1}\fi
  \endgroup}
\def\mn@eprint#1#2{\mn@eprint@#1:#2::\@nil}
\def\mn@eprint@arXiv#1{\href {http://arxiv.org/abs/#1} {{\tt arXiv:#1}}}
\def\mn@eprint@dblp#1{\href {http://dblp.uni-trier.de/rec/bibtex/#1.xml}
  {dblp:#1}}
\def\mn@eprint@#1:#2:#3:#4\@nil{\def\@tempa {#1}\def\@tempb {#2}\def\@tempc
  {#3}\ifx \@tempc \@empty \let \@tempc \@tempb \let \@tempb \@tempa \fi \ifx
  \@tempb \@empty \def\@tempb {arXiv}\fi \@ifundefined
  {mn@eprint@\@tempb}{\@tempb:\@tempc}{\expandafter \expandafter \csname
  mn@eprint@\@tempb\endcsname \expandafter{\@tempc}}}

\bibitem[\protect\citeauthoryear{Abruzzo, Narayanan, Dav{\'{e}}  \&
  Thompson}{Abruzzo et~al.}{2018}]{Abruzzo_2018}
Abruzzo M.~W.,  Narayanan D.,  Dav{\'{e}} R.,   Thompson R.,  2018

\bibitem[\protect\citeauthoryear{Agarwal, Dav{\'{e}}  \& Bassett}{Agarwal
  et~al.}{2018}]{Agarwal_2018}
Agarwal S.,  Dav{\'{e}} R.,   Bassett B.~A.,  2018, \mn@doi [MNRAS]
  {10.1093/MNRAS/STY1169}, 478, 3410

\bibitem[\protect\citeauthoryear{Alexander \& Hickox}{Alexander \&
  Hickox}{2012}]{Alexander_2012}
Alexander D.~M.,  Hickox R.~C.,  2012, \mn@doi [New Astron.]
  {10.1016/j.newar.2011.11.003}, 56, 93

\bibitem[\protect\citeauthoryear{Almaini et~al.,}{Almaini
  et~al.}{2017}]{Almaini_2017}
Almaini O.,  et~al., 2017, \mn@doi [MNRAS] {10.1093/mnras/stx1957}, 472, 1401

\bibitem[\protect\citeauthoryear{Angl{\'{e}}s-Alc{\'{a}}zar, {\"{O}}zel  \&
  Dav{\'{e}}}{Angl{\'{e}}s-Alc{\'{a}}zar et~al.}{2013}]{Angles-Alcazar_2013}
Angl{\'{e}}s-Alc{\'{a}}zar D.,  {\"{O}}zel F.,   Dav{\'{e}} R.,  2013, \mn@doi
  [ApJ] {10.1088/0004-637X/770/1/5}, 770, 5

\bibitem[\protect\citeauthoryear{Angl{\'{e}}s-Alc{\'{a}}zar, {\"{O}}zel,
  Dav{\'{e}}, Katz, Kollmeier  \& Oppenheimer}{Angl{\'{e}}s-Alc{\'{a}}zar
  et~al.}{2015}]{Angles-Alcazar_2015}
Angl{\'{e}}s-Alc{\'{a}}zar D.,  {\"{O}}zel F.,  Dav{\'{e}} R.,  Katz N.,
  Kollmeier J.~A.,   Oppenheimer B.~D.,  2015, \mn@doi [ApJ]
  {10.1088/0004-637X/800/2/127}, 800, 127

\bibitem[\protect\citeauthoryear{Angl{\'{e}}s-Alc{\'{a}}zar, Dav{\'{e}},
  Faucher-Gigu{\`{e}}re, {\"{O}}zel  \& Hopkins}{Angl{\'{e}}s-Alc{\'{a}}zar
  et~al.}{2017a}]{Angles-Alcazar_2017a}
Angl{\'{e}}s-Alc{\'{a}}zar D.,  Dav{\'{e}} R.,  Faucher-Gigu{\`{e}}re C.-A.,
  {\"{O}}zel F.,   Hopkins P.~F.,  2017a, \mn@doi [MNRAS]
  {10.1093/mnras/stw2565}, 464, 2840

\bibitem[\protect\citeauthoryear{Angl{\'{e}}s-Alc{\'{a}}zar,
  Faucher-Gigu{\`{e}}re, Kere{\v{s}}, Hopkins, Quataert  \&
  Murray}{Angl{\'{e}}s-Alc{\'{a}}zar et~al.}{2017b}]{Angles-Alcazar_2017b}
Angl{\'{e}}s-Alc{\'{a}}zar D.,  Faucher-Gigu{\`{e}}re C.~A.,  Kere{\v{s}} D.,
  Hopkins P.~F.,  Quataert E.,   Murray N.,  2017b, \mn@doi [MNRAS]
  {10.1093/mnras/stx1517}, 470, 4698

\bibitem[\protect\citeauthoryear{Aragon-Salamanca, Milvang-Jensen, Balcells,
  Merrifield, Bamford  \& Rodriguez Del~Pino}{Aragon-Salamanca
  et~al.}{2013}]{Aragon_Salamanca_2013}
Aragon-Salamanca A.,  Milvang-Jensen B.,  Balcells M.,  Merrifield M.~R.,
  Bamford S.~P.,   Rodriguez Del~Pino B.,  2013, \mn@doi [MNRAS]
  {10.1093/mnras/stt2202}, 438, 1038

\bibitem[\protect\citeauthoryear{Arnouts et~al.,}{Arnouts
  et~al.}{2007}]{Arnouts_2007}
Arnouts S.,  et~al., 2007, \mn@doi [A{\&}A] {10.1051/0004-6361:20077632}, 476,
  137

\bibitem[\protect\citeauthoryear{Baldry, Glazebrook, Brinkmann, Ivezi{\'{c}},
  Lupton, Nichol  \& Szalay}{Baldry et~al.}{2004}]{Baldry_2004}
Baldry I.~K.,  Glazebrook K.,  Brinkmann J.,  Ivezi{\'{c}} Å.,  Lupton R.~H.,
  Nichol R.~C.,   Szalay A.~S.,  2004, \mn@doi [ApJ] {10.1086/380092}, 600, 681

\bibitem[\protect\citeauthoryear{Bell et~al.,}{Bell et~al.}{2004}]{Bell_2004}
Bell E.~F.,  et~al., 2004, \mn@doi [ApJ] {10.1086/420778}, 608, 752

\bibitem[\protect\citeauthoryear{Benson}{Benson}{2010}]{Benson_2010}
Benson A.~J.,  2010, \mn@doi [Phys Rep] {10.1016/j.physrep.2010.06.001}, 495,
  33

\bibitem[\protect\citeauthoryear{Bertone \& Conselice}{Bertone \&
  Conselice}{2009}]{Bertone_2009}
Bertone S.,  Conselice C.~J.,  2009, \mn@doi [MNRAS]
  {10.1111/j.1365-2966.2009.14916.x}, 396, 2345

\bibitem[\protect\citeauthoryear{Bluck, Conselice, Buitrago, Gr{\"{u}}tzbauch,
  Hoyos, Mortlock  \& Bauer}{Bluck et~al.}{2012}]{Bluck_2012}
Bluck A. F.~L.,  Conselice C.~J.,  Buitrago F.~o.,  Gr{\"{u}}tzbauch R.,  Hoyos
  C.,  Mortlock A.,   Bauer A.~E.,  2012, \mn@doi [ApJ]
  {10.1088/0004-637X/747/1/34}, 747, 34

\bibitem[\protect\citeauthoryear{Bondi}{Bondi}{1952}]{Bondi_1952}
Bondi H.,  1952, \mn@doi [MNRAS] {10.1093/mnras/112.2.195}, 112, 195

\bibitem[\protect\citeauthoryear{Bower, Benson, Malbon, Helly, Frenk, Baugh,
  Cole  \& Lacey}{Bower et~al.}{2006}]{Bower_2006}
Bower R.~G.,  Benson A.~J.,  Malbon R.,  Helly J.~C.,  Frenk C.~S.,  Baugh
  C.~M.,  Cole S.,   Lacey C.~G.,  2006, \mn@doi [MNRAS]
  {10.1111/j.1365-2966.2006.10519.x}, 370, 645

\bibitem[\protect\citeauthoryear{Cox, Jonsson, Somerville, Primack  \&
  Dekel}{Cox et~al.}{2008}]{Cox_2007}
Cox T.~J.,  Jonsson P.,  Somerville R.~S.,  Primack J.~R.,   Dekel A.,  2008,
  \mn@doi [MNRAS] {10.1111/j.1365-2966.2007.12730.x}, 384, 386

\bibitem[\protect\citeauthoryear{Croton et~al.,}{Croton
  et~al.}{2006}]{Croton_2006}
Croton D.~J.,  et~al., 2006, \mn@doi [MNRAS]
  {10.1111/j.1365-2966.2005.09675.x}, 365, 11

\bibitem[\protect\citeauthoryear{Dav{\'{e}}, Finlator  \&
  Oppenheimer}{Dav{\'{e}} et~al.}{2012}]{Dave_2012}
Dav{\'{e}} R.,  Finlator K.,   Oppenheimer B.~D.,  2012, \mn@doi [MNRAS]
  {10.1111/j.1365-2966.2011.20148.x}, 421, 98

\bibitem[\protect\citeauthoryear{Dav{\'{e}}, Thompson  \& Hopkins}{Dav{\'{e}}
  et~al.}{2016}]{Dave_2016}
Dav{\'{e}} R.,  Thompson R.,   Hopkins P.~F.,  2016, \mn@doi [MNRAS]
  {10.1093/mnras/stw1862}, 462, 3265

\bibitem[\protect\citeauthoryear{Dav{\'{e}}, Rafieferantsoa  \&
  Thompson}{Dav{\'{e}} et~al.}{2017}]{Dave_2017}
Dav{\'{e}} R.,  Rafieferantsoa M.~H.,   Thompson R.~J.,  2017, \mn@doi [MNRAS]
  {10.1093/MNRAS/STX1693}, 471, 1671

\bibitem[\protect\citeauthoryear{Dav{\'{e}}, Angl{\'{e}}s-Alc{\'{a}}zar,
  Narayanan, Li, Rafieferantsoa  \& Appleby}{Dav{\'{e}}
  et~al.}{2019}]{Dave_2019}
Dav{\'{e}} R.,  Angl{\'{e}}s-Alc{\'{a}}zar D.,  Narayanan D.,  Li Q.,
  Rafieferantsoa M.~H.,   Appleby S.,  2019, \mn@doi [MNRAS]
  {10.1093/mnras/stz937}, 486, 2827

\bibitem[\protect\citeauthoryear{Dekel, Sari  \& Ceverino}{Dekel
  et~al.}{2009}]{Dekel_2009}
Dekel A.,  Sari R.,   Ceverino D.,  2009, \mn@doi [ApJ]
  {10.1088/0004-637X/703/1/785}, 703, 785

\bibitem[\protect\citeauthoryear{Di~Matteo, Springel  \& Hernquist}{Di~Matteo
  et~al.}{2005}]{DiMatteo_2005}
Di~Matteo T.,  Springel V.,   Hernquist L.,  2005, \mn@doi [Nat]
  {10.1038/nature03335}, 433, 604

\bibitem[\protect\citeauthoryear{Diamond-Stanic, Moustakas, Tremonti, Coil,
  Hickox, Robaina, Rudnick  \& Sell}{Diamond-Stanic
  et~al.}{2012}]{Diamond_Stanic_2012}
Diamond-Stanic A.~M.,  Moustakas J.,  Tremonti C.~A.,  Coil A.~L.,  Hickox
  R.~C.,  Robaina A.~R.,  Rudnick G.~H.,   Sell P.~H.,  2012, \mn@doi [ApJL]
  {10.1088/2041-8205/755/2/L26}, 755, L26

\bibitem[\protect\citeauthoryear{Dierckx}{Dierckx}{1975}]{Dierckx_1975}
Dierckx P.,  1975, \mn@doi [J Comput Appl Math]
  {https://doi.org/10.1016/0771-050X(75)90034-0}, 1, 165

\bibitem[\protect\citeauthoryear{Duncan et~al.,}{Duncan
  et~al.}{2019}]{Duncan_2019}
Duncan K.,  et~al., 2019, \mn@doi [ApJ] {10.3847/1538-4357/ab148a}, 876, 110

\bibitem[\protect\citeauthoryear{Ellison, Patton, Simard  \&
  McConnachie}{Ellison et~al.}{2008}]{Ellison_2008}
Ellison S.~L.,  Patton D.~R.,  Simard L.,   McConnachie A.~W.,  2008, \mn@doi
  [Astronomical Journal] {10.1088/0004-6256/135/5/1877}, 135, 1877

\bibitem[\protect\citeauthoryear{Ellison, Patton, Mendel  \& Scudder}{Ellison
  et~al.}{2011}]{Ellison_2011}
Ellison S.~L.,  Patton D.~R.,  Mendel J.~T.,   Scudder J.~M.,  2011, \mn@doi
  [MNRAS] {10.1111/j.1365-2966.2011.19624.x}, 418, 2043

\bibitem[\protect\citeauthoryear{Ellison, Viswanathan, Patton, Bottrell,
  McConnachie, Gwyn  \& Cuillandre}{Ellison et~al.}{2019}]{Ellison_2019}
Ellison S.~L.,  Viswanathan A.,  Patton D.~R.,  Bottrell C.,  McConnachie
  A.~W.,  Gwyn S.,   Cuillandre J.-C.,  2019, \mn@doi [MNRAS]
  {10.1093/mnras/stz1431}, 487, 2491

\bibitem[\protect\citeauthoryear{Faber et~al.,}{Faber
  et~al.}{2007}]{Faber_2007}
Faber S.~M.,  et~al., 2007, \mn@doi [ApJ] {10.1086/519294}, 665, 265

\bibitem[\protect\citeauthoryear{F{\"{o}}rster~Schreiber
  et~al.,}{F{\"{o}}rster~Schreiber et~al.}{2009}]{Forster-Schreiber_2009}
F{\"{o}}rster~Schreiber N.~M.,  et~al., 2009, \mn@doi [ApJ]
  {10.1088/0004-637X/706/2/1364}, 706, 1364

\bibitem[\protect\citeauthoryear{Gabor \& Dav{\'{e}}}{Gabor \&
  Dav{\'{e}}}{2012}]{Gabor_2012}
Gabor J.~M.,  Dav{\'{e}} R.,  2012, \mn@doi [MNRAS]
  {10.1111/j.1365-2966.2012.21640.x}, 427, 1816

\bibitem[\protect\citeauthoryear{Gabor \& Dav{\'{e}}}{Gabor \&
  Dav{\'{e}}}{2015}]{Gabor_2015}
Gabor J.~M.,  Dav{\'{e}} R.,  2015, \mn@doi [MNRAS] {10.1093/mnras/stu2399},
  447, 374

\bibitem[\protect\citeauthoryear{Gabor, Dav{\'{e}}, Finlator  \&
  Oppenheimer}{Gabor et~al.}{2010}]{Gabor_2010}
Gabor J.~M.,  Dav{\'{e}} R.,  Finlator K.,   Oppenheimer B.~D.,  2010, \mn@doi
  [MNRAS] {10.1111/j.1365-2966.2010.16961.x}, 407, 749

\bibitem[\protect\citeauthoryear{Genel, Genzel, Bouch{\'{e}}, Naab  \&
  Sternberg}{Genel et~al.}{2009}]{Genel_2009}
Genel S.,  Genzel R.,  Bouch{\'{e}} N.,  Naab T.,   Sternberg A.,  2009,
  \mn@doi [ApJ] {10.1088/0004-637X/701/2/2002}, 701, 2002

\bibitem[\protect\citeauthoryear{Heckman \& Best}{Heckman \&
  Best}{2014}]{Heckman_2014}
Heckman T.,  Best P.,  2014, \mn@doi [ARA{\&}A]
  {10.1146/annurev-astro-081913-035722}, 52, 589

\bibitem[\protect\citeauthoryear{Henden, Puchwein, Shen  \& Sijacki}{Henden
  et~al.}{2018}]{Henden_2018}
Henden N.~A.,  Puchwein E.,  Shen S.,   Sijacki D.,  2018, \mn@doi [MNRAS]
  {10.1093/mnras/sty1780}, 479, 5385

\bibitem[\protect\citeauthoryear{Hewlett, Villforth, Wild, Mendez-Abreu, Pawlik
   \& Rowlands}{Hewlett et~al.}{2017}]{Hewlett_2017}
Hewlett T.,  Villforth C.,  Wild V.,  Mendez-Abreu J.,  Pawlik M.,   Rowlands
  K.,  2017, \mn@doi [MNRAs] {10.1093/mnras/stx997}, 470, 755

\bibitem[\protect\citeauthoryear{Hopkins}{Hopkins}{2015}]{Hopkins_2015}
Hopkins P.~F.,  2015, \mn@doi [MNRAS] {10.1093/mnras/stv195}, 450, 53

\bibitem[\protect\citeauthoryear{Hopkins, Hernquist, Cox, Di~Matteo, Robertson
  \& Springel}{Hopkins et~al.}{2005}]{Hopkins_2006}
Hopkins P.~F.,  Hernquist L.,  Cox T.~J.,  Di~Matteo T.,  Robertson B.,
  Springel V.,  2005, \mn@doi [ApJ] {10.1086/499298}, 163, 1

\bibitem[\protect\citeauthoryear{Hopkins, Hernquist, Cox  \&
  Kere{\v{s}}}{Hopkins et~al.}{2008}]{Hopkins_2008}
Hopkins P.~F.,  Hernquist L.,  Cox T.~J.,   Kere{\v{s}} D.,  2008, \mn@doi
  [ApJS] {10.1086/524362}, 175, 356

\bibitem[\protect\citeauthoryear{Hopkins, Cox, Younger  \& Hernquist}{Hopkins
  et~al.}{2009}]{Hopkins_2009}
Hopkins P.~F.,  Cox T.~J.,  Younger J.~D.,   Hernquist L.,  2009, \mn@doi [ApJ]
  {10.1088/0004-637X/691/2/1168}, 691, 1168

\bibitem[\protect\citeauthoryear{Hopkins, Quataert  \& Murray}{Hopkins
  et~al.}{2011}]{Hopkins_2011}
Hopkins P.~F.,  Quataert E.,   Murray N.,  2011, \mn@doi [MNRAS]
  {10.1111/j.1365-2966.2011.19306.x}, 417, 950

\bibitem[\protect\citeauthoryear{Ilbert et~al.,}{Ilbert
  et~al.}{2013}]{Ilbert_2013}
Ilbert O.,  et~al., 2013, \mn@doi [A{\&}A] {10.1051/0004-6361/201321100}, 556,
  A55

\bibitem[\protect\citeauthoryear{Jogee et~al.,}{Jogee
  et~al.}{2009}]{Jogee_2009}
Jogee S.,  et~al., 2009, \mn@doi [ApJ] {10.1088/0004-637X/697/2/1971}, 697,
  1971

\bibitem[\protect\citeauthoryear{Johansson, Naab  \& Burkert}{Johansson
  et~al.}{2009}]{Johansson_2009}
Johansson P.~H.,  Naab T.,   Burkert A.,  2009, \mn@doi [ApJ]
  {10.1088/0004-637X/690/1/802}, 690, 802

\bibitem[\protect\citeauthoryear{Kassin et~al.,}{Kassin
  et~al.}{2012}]{Kassin_2012}
Kassin S.~A.,  et~al., 2012, \mn@doi [ApJ] {10.1088/0004-637X/758/2/106}, 758,
  106

\bibitem[\protect\citeauthoryear{Kaviraj}{Kaviraj}{2014}]{Kaviraj_2014}
Kaviraj S.,  2014, \mn@doi [MNRAS] {10.1093/mnrasl/slt136}, 437, L41

\bibitem[\protect\citeauthoryear{Kocevski et~al.,}{Kocevski
  et~al.}{2012}]{Kocevski_2012}
Kocevski D.~D.,  et~al., 2012, \mn@doi [ApJ] {10.1088/0004-637X/744/2/148}, 744

\bibitem[\protect\citeauthoryear{Konar, Hardcastle, Jamrozy, Croston  \&
  Nandi}{Konar et~al.}{2012}]{Konar_2012}
Konar C.,  Hardcastle M.~J.,  Jamrozy M.,  Croston J.~H.,   Nandi S.,  2012,
  \mn@doi [MNRAS] {10.1111/j.1365-2966.2012.21279.x}, 424, 1061

\bibitem[\protect\citeauthoryear{Krumholz, McKee  \& Tumlinson}{Krumholz
  et~al.}{2009}]{Krumholz_2009}
Krumholz M.~R.,  McKee C.~F.,   Tumlinson J.,  2009, \mn@doi [ApJ]
  {10.1088/0004-637x/699/1/850}, 699, 850

\bibitem[\protect\citeauthoryear{Kurczynski et~al.,}{Kurczynski
  et~al.}{2016}]{Kurczynski_2016}
Kurczynski P.,  et~al., 2016, \mn@doi [ApJ] {10.3847/2041-8205/820/1/L1}, 820,
  L1

\bibitem[\protect\citeauthoryear{Leja, Carnall, Johnson, Conroy  \&
  Speagle}{Leja et~al.}{2018}]{Leja_2018}
Leja J.,  Carnall A.~C.,  Johnson B.~D.,  Conroy C.,   Speagle J.~S.,  2018, ]
  {10.3847/1538-4357/ab133c}

\bibitem[\protect\citeauthoryear{Li, Narayanan  \& Dav{\'{e}}}{Li
  et~al.}{2019}]{Li_2019}
Li Q.,  Narayanan D.,   Dav{\'{e}} R.,  2019

\bibitem[\protect\citeauthoryear{Liske et~al.,}{Liske
  et~al.}{2015}]{Liske_2015}
Liske J.,  et~al., 2015, \mn@doi [MNRAS] {10.1093/mnras/stv1436}, 452, 2087

\bibitem[\protect\citeauthoryear{Lofthouse, Kaviraj, Conselice, Mortlock  \&
  Hartley}{Lofthouse et~al.}{2017}]{Lofthouse_2017}
Lofthouse E.~K.,  Kaviraj S.,  Conselice C.~J.,  Mortlock A.,   Hartley W.,
  2017, \mn@doi [MNRAS] {10.1093/mnras/stw2895}, 465, 2895

\bibitem[\protect\citeauthoryear{Lotz, Jonsson, Cox, Croton, Primack,
  Somerville  \& Stewart}{Lotz et~al.}{2011}]{Lotz_2011}
Lotz J.~M.,  Jonsson P.,  Cox T.~J.,  Croton D.,  Primack J.~R.,  Somerville
  R.~S.,   Stewart K.,  2011, \mn@doi [ApJ] {10.1088/0004-637X/742/2/103}, 742,
  103

\bibitem[\protect\citeauthoryear{Mahajan}{Mahajan}{2013}]{Mahajan_2013}
Mahajan S.,  2013, \mn@doi [MNRAS] {10.1093/mnrasl/slt021}, 431, L117

\bibitem[\protect\citeauthoryear{Maltby, Almaini, Wild, Hatch, Hartley,
  Simpson, Rowlands  \& Socolovsky}{Maltby et~al.}{2018}]{Maltby_2018}
Maltby D.~T.,  Almaini O.,  Wild V.,  Hatch N.~A.,  Hartley W.~G.,  Simpson C.,
   Rowlands K.,   Socolovsky M.,  2018, \mn@doi [MNRAS]
  {10.1093/mnras/sty1794}, 480, 381

\bibitem[\protect\citeauthoryear{Mancini et~al.,}{Mancini
  et~al.}{2019}]{Mancini_2019}
Mancini C.,  et~al., 2019

\bibitem[\protect\citeauthoryear{Martig, Bournaud, Teyssier  \& Dekel}{Martig
  et~al.}{2009}]{Martig_2009}
Martig M.,  Bournaud F.,  Teyssier R.,   Dekel A.,  2009, \mn@doi [ApJ]
  {10.1088/0004-637x/707/1/250}, 707, 250

\bibitem[\protect\citeauthoryear{Martin, Kaviraj, Devriendt, Dubois  \&
  Pichon}{Martin et~al.}{2018}]{Martin_2018}
Martin G.,  Kaviraj S.,  Devriendt J.~E.~G.,  Dubois Y.,   Pichon C.,  2018,
  \mn@doi [MNRAS] {10.1093/mnras/sty1936}, 480, 2266

\bibitem[\protect\citeauthoryear{McNamara \& Nulsen}{McNamara \&
  Nulsen}{2007}]{McNamara_2007}
McNamara B.~R.,  Nulsen P.~E.~J.,  2007, \mn@doi [ARA{\&}A]
  {10.1146/annurev.astro.45.051806.110625}, 45, 117

\bibitem[\protect\citeauthoryear{Mihos \& Hernquist}{Mihos \&
  Hernquist}{1996}]{Mihos_1996}
Mihos J.~C.,  Hernquist L.,  1996, \mn@doi [ApJ] {10.1086/177353}, 464, 641

\bibitem[\protect\citeauthoryear{Moreno et~al.,}{Moreno
  et~al.}{2019}]{Moreno_2019}
Moreno J.,  et~al., 2019, \mn@doi [MNRAS] {10.1093/mnras/stz417}, 485, 1320

\bibitem[\protect\citeauthoryear{Muzzin et~al.,}{Muzzin
  et~al.}{2013}]{Muzzin_2013}
Muzzin A.,  et~al., 2013, \mn@doi [ApJ] {10.1088/0004-637X/777/1/18}, 777, 18

\bibitem[\protect\citeauthoryear{Naab \& Ostriker}{Naab \&
  Ostriker}{2017}]{Naab_Ostriker_2017}
Naab T.,  Ostriker J.~P.,  2017, \mn@doi [ARA{\&}A]
  {10.1146/annurev-astro-081913-040019}, 55, 59

\bibitem[\protect\citeauthoryear{Noeske et~al.,}{Noeske
  et~al.}{2007}]{Noeske_2007}
Noeske K.~G.,  et~al., 2007, \mn@doi [ApJL] {10.1086/517926}, 660, L43

\bibitem[\protect\citeauthoryear{Pacifici et~al.,}{Pacifici
  et~al.}{2016}]{Pacifici_2016}
Pacifici C.,  et~al., 2016, \mn@doi [ApJ] {10.3847/0004-637x/832/1/79}, 832, 79

\bibitem[\protect\citeauthoryear{Pawlik, Wild, Walcher, Johansson, Villforth,
  Rowlands, Mendez-Abreu  \& Hewlett}{Pawlik et~al.}{2016}]{Pawlik_2016}
Pawlik M.~M.,  Wild V.,  Walcher C.~J.,  Johansson P.~H.,  Villforth C.,
  Rowlands K.,  Mendez-Abreu J.,   Hewlett T.,  2016, \mn@doi [MNRAS]
  {10.1093/mnras/stv2878}, 456, 3032

\bibitem[\protect\citeauthoryear{Pawlik et~al.,}{Pawlik
  et~al.}{2018}]{Pawlik_2018}
Pawlik M.~M.,  et~al., 2018, \mn@doi [MNRAS] {10.1093/mnras/sty589}, 477, 1708

\bibitem[\protect\citeauthoryear{Pawlik, McAlpine, Trayford, Wild, Bower,
  Crain, Schaller  \& Schaye}{Pawlik et~al.}{2019}]{Pawlik_2019}
Pawlik M.~M.,  McAlpine S.,  Trayford J.~W.,  Wild V.,  Bower R.,  Crain R.~A.,
   Schaller M.,   Schaye J.,  2019, \mn@doi [Nat.] {10.1038/s41550-019-0725-z},
  3, 440

\bibitem[\protect\citeauthoryear{Peacock}{Peacock}{1983}]{Peacock_1983}
Peacock J.~A.,  1983, \mn@doi [MNRAS] {10.1093/mnras/202.3.615}, 202, 615

\bibitem[\protect\citeauthoryear{Peng et~al.,}{Peng et~al.}{2010}]{Peng_2010}
Peng Y.-j.,  et~al., 2010, \mn@doi [ApJ] {10.1088/0004-637X/721/1/193}, 721,
  193

\bibitem[\protect\citeauthoryear{Perna, Lanzuisi, Brusa, Mignoli  \&
  Cresci}{Perna et~al.}{2017}]{Perna_2017}
Perna M.,  Lanzuisi G.,  Brusa M.,  Mignoli M.,   Cresci G.,  2017, \mn@doi
  [A{\&}A] {10.1051/0004-6361/201630369}, 603, A99

\bibitem[\protect\citeauthoryear{Pillepich et~al.,}{Pillepich
  et~al.}{2018}]{Pillepich_2018}
Pillepich A.,  et~al., 2018, \mn@doi [MNRAS] {10.1093/mnras/stx3112}, 475, 648

\bibitem[\protect\citeauthoryear{{Planck Collaboration} et~al.,}{{Planck
  Collaboration} et~al.}{2016}]{Planck_2016}
{Planck Collaboration} et~al., 2016, \mn@doi [A{\&}A]
  {10.1051/0004-6361/201525830}, 594, A13

\bibitem[\protect\citeauthoryear{Poggianti et~al.,}{Poggianti
  et~al.}{2009}]{Poggianti_2009}
Poggianti B.~M.,  et~al., 2009, \mn@doi [ApJ] {10.1088/0004-637X/697/2/L137},
  697

\bibitem[\protect\citeauthoryear{Rafieferantsoa, Dav{\'{e}}  \&
  Naab}{Rafieferantsoa et~al.}{2018}]{Rafieferantsoa_2019}
Rafieferantsoa M.,  Dav{\'{e}} R.,   Naab T.,  2018, \mn@doi [MNRAS]
  {10.1093/mnras/stz1199}, 486, 5184

\bibitem[\protect\citeauthoryear{Rees \& Ostriker}{Rees \&
  Ostriker}{1977}]{Rees_1977}
Rees M.~J.,  Ostriker J.~P.,  1977, \mn@doi [MNRAS] {10.1093/mnras/179.4.541},
  179, 541

\bibitem[\protect\citeauthoryear{Robertson, Bullock, Cox, Di~Matteo, Hernquist,
  Springel  \& Yoshida}{Robertson et~al.}{2006}]{Robertson_2006}
Robertson B.,  Bullock J.~S.,  Cox T.~J.,  Di~Matteo T.,  Hernquist L.,
  Springel V.,   Yoshida N.,  2006, \mn@doi [ApJ] {10.1086/504412}, 645, 986

\bibitem[\protect\citeauthoryear{Rodighiero et~al.,}{Rodighiero
  et~al.}{2011}]{Rodighiero_2011}
Rodighiero G.,  et~al., 2011, \mn@doi [ApJ] {10.1088/2041-8205/739/2/l40}, 739,
  L40

\bibitem[\protect\citeauthoryear{Saikia, Gupta  \& Konar}{Saikia
  et~al.}{2007}]{Saikia_2007}
Saikia D.~J.,  Gupta N.,   Konar C.,  2007, \mn@doi [MNRAS]
  {10.1111/j.1745-3933.2006.00269.x}, 375, L31

\bibitem[\protect\citeauthoryear{Sanders \& Mirabel}{Sanders \&
  Mirabel}{1996}]{Sanders_Mirabel_1996}
Sanders D.~B.,  Mirabel I.~F.,  1996, \mn@doi [ARA{\&}A]
  {10.1146/annurev.astro.34.1.749}, 34, 749

\bibitem[\protect\citeauthoryear{Schawinski, Treister, Urry, Cardamone, Simmons
   \& Yi}{Schawinski et~al.}{2011}]{Schawinski_2011}
Schawinski K.,  Treister E.,  Urry C.~M.,  Cardamone C.~N.,  Simmons B.,   Yi
  S.~K.,  2011, \mn@doi [ApJL] {10.1088/2041-8205/727/2/L31}, 727

\bibitem[\protect\citeauthoryear{Schawinski et~al.,}{Schawinski
  et~al.}{2014}]{Schawinski_2014}
Schawinski K.,  et~al., 2014, \mn@doi [MNRAS] {10.1093/mnras/stu327}, 440, 889

\bibitem[\protect\citeauthoryear{Schaye et~al.,}{Schaye
  et~al.}{2015}]{Schaye_2015}
Schaye J.,  et~al., 2015, \mn@doi [MNRAS] {10.1093/mnras/stu2058}, 446, 521

\bibitem[\protect\citeauthoryear{Schmidt}{Schmidt}{1959}]{Schmidt_1959}
Schmidt M.,  1959, \mn@doi [ApJ] {10.1086/146614}, 129, 243

\bibitem[\protect\citeauthoryear{Schreiber et~al.,}{Schreiber
  et~al.}{2018}]{Schreiber_2018}
Schreiber C.,  et~al., 2018, \mn@doi [A{\&}A] {10.1051/0004-6361/201833070},
  618, A85

\bibitem[\protect\citeauthoryear{Sijacki, Springel, Di~Matteo  \&
  Hernquist}{Sijacki et~al.}{2007}]{Sijacki_2007}
Sijacki D.,  Springel V.,  Di~Matteo T.,   Hernquist L.,  2007, \mn@doi [MNRAS]
  {10.1111/j.1365-2966.2007.12153.x}, 380, 877

\bibitem[\protect\citeauthoryear{Smith et~al.,}{Smith
  et~al.}{2016}]{Smith_2018}
Smith B.~D.,  et~al., 2016, \mn@doi [MNRAS] {10.1093/mnras/stw3291}, 466, 2217

\bibitem[\protect\citeauthoryear{Socolovsky, Maltby, Hatch, Almaini, Wild,
  Hartley, Simpson  \& Rowlands}{Socolovsky et~al.}{2019}]{Socolovsky_2018}
Socolovsky M.,  Maltby D.~T.,  Hatch N.~A.,  Almaini O.,  Wild V.,  Hartley
  W.~G.,  Simpson C.,   Rowlands K.,  2019, \mn@doi [MNRAS]
  {10.1093/mnras/sty2840}, 482, 1640

\bibitem[\protect\citeauthoryear{Somerville \& Dav{\'{e}}}{Somerville \&
  Dav{\'{e}}}{2015}]{Somerville_2015}
Somerville R.~S.,  Dav{\'{e}} R.,  2015, \mn@doi [ARA{\&}A]
  {10.1146/annurev-astro-082812-140951}, 53, 51

\bibitem[\protect\citeauthoryear{Somerville, Hopkins, Cox, Robertson  \&
  Hernquist}{Somerville et~al.}{2008}]{Somerville_2008}
Somerville R.~S.,  Hopkins P.~F.,  Cox T.~J.,  Robertson B.~E.,   Hernquist L.,
   2008, \mn@doi [MNRAS] {10.1111/j.1365-2966.2008.13805.x}, 391, 481

\bibitem[\protect\citeauthoryear{Speagle, Steinhardt, Capak  \&
  Silverman}{Speagle et~al.}{2014}]{Speagle_2014}
Speagle J.~S.,  Steinhardt C.~L.,  Capak P.~L.,   Silverman J.~D.,  2014,
  \mn@doi [ApJS] {10.1088/0067-0049/214/2/15}, 214, 15

\bibitem[\protect\citeauthoryear{Springel}{Springel}{2005}]{Springel_2005}
Springel V.,  2005, \mn@doi [MNRAS] {10.1111/j.1365-2966.2005.09655.x}, 364,
  1105

\bibitem[\protect\citeauthoryear{Springel et~al.,}{Springel
  et~al.}{2005}]{Springel_2005b}
Springel V.,  et~al., 2005, \mn@doi [Nat] {10.1038/nature03597}, 435, 629

\bibitem[\protect\citeauthoryear{Sturm et~al.,}{Sturm
  et~al.}{2011}]{Sturm_2011}
Sturm E.,  et~al., 2011, \mn@doi [ApJL] {10.1088/2041-8205/733/1/L16}, 733, L16

\bibitem[\protect\citeauthoryear{Tacchella, Dekel, Carollo, Ceverino, DeGraf,
  Lapiner, Mandelker  \& Primack}{Tacchella et~al.}{2016}]{Tacchella_2016}
Tacchella S.,  Dekel A.,  Carollo C.~M.,  Ceverino D.,  DeGraf C.,  Lapiner S.,
   Mandelker N.,   Primack J.~R.,  2016, \mn@doi [MNRAS]
  {10.1093/mnras/stw131}, 457, 2790

\bibitem[\protect\citeauthoryear{Thomas, Dav{\'{e}}, Angl{\'{e}}s-Alc{\'{a}}zar
   \& Jarvis}{Thomas et~al.}{2019}]{Thomas_2019}
Thomas N.,  Dav{\'{e}} R.,  Angl{\'{e}}s-Alc{\'{a}}zar D.,   Jarvis M.,  2019

\bibitem[\protect\citeauthoryear{Villforth et~al.,}{Villforth
  et~al.}{2014}]{Villforth_2014}
Villforth C.,  et~al., 2014, \mn@doi [MNRAs] {10.1093/mnras/stu173}, 439, 3342

\bibitem[\protect\citeauthoryear{Vogelsberger et~al.,}{Vogelsberger
  et~al.}{2014}]{Vogelsberger_2014}
Vogelsberger M.,  et~al., 2014, \mn@doi [MNRAS] {10.1093/mnras/stu1536}, 444,
  1518

\bibitem[\protect\citeauthoryear{White \& Frenk}{White \&
  Frenk}{2002}]{WhiteFrenk_1991}
White S. D.~M.,  Frenk C.~S.,  2002, \mn@doi [ApJ] {10.1086/170483}, 379, 52

\bibitem[\protect\citeauthoryear{Wild, Walcher, Johansson, Tresse, Charlot,
  Pollo, Le~F{\`{e}}vre  \& de Ravel}{Wild et~al.}{2009}]{Wild_2009}
Wild V.,  Walcher C.~J.,  Johansson P.~H.,  Tresse L.,  Charlot S.,  Pollo A.,
  Le~F{\`{e}}vre O.,   de Ravel L.,  2009, \mn@doi [MNRAS]
  {10.1111/j.1365-2966.2009.14537.x}, 395, 144

\bibitem[\protect\citeauthoryear{Wild, Heckman  \& Charlot}{Wild
  et~al.}{2010}]{Wild_2010}
Wild V.,  Heckman T.,   Charlot S.,  2010, \mn@doi [MNRAS]
  {10.1111/j.1365-2966.2010.16536.x}

\bibitem[\protect\citeauthoryear{Wild, Almaini, Dunlop, Simpson, Rowlands,
  Bowler, Maltby  \& McLure}{Wild et~al.}{2016}]{Wild_2016}
Wild V.,  Almaini O.,  Dunlop J.,  Simpson C.,  Rowlands K.,  Bowler R.,
  Maltby D.,   McLure R.,  2016, \mn@doi [MNRAS] {10.1093/mnras/stw1996}, 463,
  832

\bibitem[\protect\citeauthoryear{Wilkinson, Pimbblet  \& Stott}{Wilkinson
  et~al.}{2017}]{Wilkinson_2017}
Wilkinson C.~L.,  Pimbblet K.~A.,   Stott J.~P.,  2017, \mn@doi [MNRAS]
  {10.1093/mnras/stx2034}, 472, 1447

\bibitem[\protect\citeauthoryear{Wuyts et~al.,}{Wuyts
  et~al.}{2011}]{Wuyts_2011}
Wuyts S.,  et~al., 2011, \mn@doi [ApJ] {10.1088/0004-637X/742/2/96}, 742, 96

\bibitem[\protect\citeauthoryear{Yang, Tremonti, Zabludoff  \& Zaritsky}{Yang
  et~al.}{2006}]{Yang_2006}
Yang Y.,  Tremonti C.~A.,  Zabludoff A.~I.,   Zaritsky D.,  2006, \mn@doi [ApJ]
  {10.1086/506909}, 646, L33

\bibitem[\protect\citeauthoryear{Yuan, Kewley  \& Sanders}{Yuan
  et~al.}{2010}]{Yuan_2010}
Yuan T.~T.,  Kewley L.~J.,   Sanders D.~B.,  2010, \mn@doi [ApJ]
  {10.1088/0004-637X/709/2/884}, 709, 884

\bibitem[\protect\citeauthoryear{Zabludoff, Zaritsky, Lin, Tucker, Hashimoto,
  Shectman, Oemler  \& Kirshner}{Zabludoff et~al.}{1996}]{Zabludof_1996}
Zabludoff A.~I.,  Zaritsky D.,  Lin H.,  Tucker D.,  Hashimoto Y.,  Shectman
  S.~A.,  Oemler A.,   Kirshner R.~P.,  1996, \mn@doi [ApJ] {10.1086/177495},
  466, 104

\bibitem[\protect\citeauthoryear{Zezas, Hernquist, Fabbiano  \& Miller}{Zezas
  et~al.}{2003}]{Zezas_2003}
Zezas A.,  Hernquist L.,  Fabbiano G.,   Miller J.,  2003, \mn@doi [ApJL]
  {10.1086/380895}, 599, L73

\makeatother
\end{thebibliography}

\appendix
\section{Supporting material}
Codes for merger identification and quenching time estimator are publicly available at \url{https://github.com/Currodri/SH_Project/tree/master} .

\bsp	
\label{lastpage}
\end{document}